\documentclass[12pt,preprint]{aastex}
\usepackage{natbib,graphicx,latexsym}
\newcommand{\lsim}{\
\raise-2.truept\hbox{\rlap{\hbox{$\sim$}}\raise5.truept\hbox{$<$}\ }}
\newcommand{\gsim}{\
\raise-2.truept\hbox{\rlap{\hbox{$\sim$}}\raise5.truept\hbox{$>$}\ }}

\begin{document}
\shorttitle{Radial SBF \& color gradients in E-type galaxies with ACS}
\title{Detection of Radial Surface Brightness Fluctuation and Color Gradients 
in elliptical galaxies with ACS\altaffilmark{1}}

\author{Michele Cantiello\altaffilmark{2}, John P.~Blakeslee\altaffilmark{3}, 
Gabriella Raimondo\altaffilmark{2}, Simona Mei\altaffilmark{3}, Enzo
Brocato\altaffilmark{2}, Massimo Capaccioli\altaffilmark{4,5}}
\altaffiltext{1}{Based on observations made with the NASA/ESA Hubble Space
Telescope, which is operated by the Association of Universities for
Research in Astronomy, Inc., under NASA contract NAS 5-26555. These
observations are associated with programs \#9427, \#9293, \#9399.}  
\altaffiltext{2}{INAF--Osservatorio Astronomico di
Teramo, Via M. Maggini, I-64100 Teramo, Italy}
\altaffiltext{3}{Department of Physics and Astronomy, Johns Hopkins University,
Baltimore, MD 21218.}  
\altaffiltext{4}{Dipartimento di Scienze Fisiche, Universit\`a Federico II di
Napoli, Complesso Monte S. Angelo, via Cintia, 80126, Napoli, Italy}
\altaffiltext{5}{INAF--Osservatorio Astronomico di Capodimonte, via
Moiariello 16, 80131 Napoli, Italy}

\begin{abstract}

We study surface brightness fluctuations (SBF) in a sample of 8
elliptical galaxies using Advanced Camera for Surveys (ACS) Wide Field
Channel (WFC) data drawn from the {\it Hubble Space Telescope} (HST)
archive.  SBF magnitudes in the F814W bandpass, and galaxy colors from
F814W, F435W, and F606W images -- when available -- are
presented. Galaxy surface brightness profiles are determined as well.
We present the first SBF--broadband color calibration for the ACS/WFC
$F814W$ bandpass, and (relative) distance moduli estimates for 7 of
our galaxies.

We detect and study in detail the SBF variations within individual
galaxies as a probe of possible changes in the underlying stellar
populations.  Inspecting both the SBF and color gradients in
comparison to model predictions, we argue that SBF, and SBF-gradients,
can in principle be used for unraveling the different evolutionary
paths taken by galaxies, though a more comprehensive study of this
issue would be required.  We confirm that the radial variation of
galaxy stellar population properties should be mainly connected to the
presence of radial chemical abundance gradients, with the outer galaxy
regions being more metal poor than the inner ones.

\end{abstract}

\section{Introduction}

Stars, as the basic constituents of galaxies, mark the dynamical and
chemical evolution of galaxies.  Since a direct study of individual
stars can be performed only for the nearest extragalactic systems, to
analyze the properties of distant galaxies, one must rely on the
integrated light from all the stars along a given line of sight.
Therefore, broadband colors and line-strength measurements have been
widely used to study galaxy properties, and early-type galaxies have
been the preferred targets for integrated light studies. In contrast
to spirals, which are a complex of young, intermediate, and old stars
and patchy dust, there is general agreement on the view of ellipticals
as systems without current star formation and little dust
\citep[e.g.,][]{cimatti04}. Thus, their integrated light should be
dominated by the old stellar component.  Although this view is
continually revisited and refined, it remains the baseline against
which new observations of early-type galaxies are compared
\citep[e.g.,][]{renzini99,bernardi03}.  Furthermore, integrated
starlight studies remain the primary means for inferring information
about the physical properties of elliptical galaxies (see for example
the review by \citealt{kormendy89} and references therein) and
their possible formation histories \citep[e.g.,][]{peebles02}.

In this work we give some new photometric information useful to
analyze the stellar content of ellipticals. Although such targets are
not complicated by the presence of young stars or extensive dust, they
cannot simply be identified as ``Population II'' systems, as
originally supposed \citep{baade44}.  In the last decades the
interpretation of broadband colors and spectral indices has lead to a
revision of the earlier assumption of ellipticals as composed of a
uniform metal rich, old stellar population, pointing out not only that
different galaxies are characterized by populations with different age
and chemical composition, but also that such differences exist among
different regions of the same galaxy.

Since the work of \citet{pagel81} on the radial distribution
of metals in external galaxies, integrated spectrophotometric studies
evidenced the presence of radial gradients, typically indicating
chemical abundance variations between the central and the outer parts
in both elliptical and spiral galaxies, generally decreasing outward.

Evidence also exists that abundance gradients become flatter as one
moves from late- to early-type galaxies \citep{henry99}.
Furthermore, within ellipticals the differences in the observed
gradients is found to be only weakly correlated, if at all, with
any other physical property of the galaxy: dynamical mass, absolute
magnitude, central velocity dispersion \citep{kobayashi04}. The merging
formation scenario could explain such behaviors, with the flatter gradients 
resulting from more complex merging histories \citep{white80}.

However, broadband colors and line-strength analyses are only partly
sufficient to perform a detailed reconstruction of the galaxy
formation and evolutionary history, due to the confusing effects of
age and metallicity. As well known, the analysis of integrated light
is subject to an age/metallicity degeneracy.  Most of the stellar
population synthesis models, predict a $\Delta log Age / \Delta log [Fe/H]
\sim -3/2$, which means that a change of a factor of $\sim\,$3 in the age can
mimic the effects of a change of a factor of $\sim\,$2 in metallicity
[Fe/H] \citep{worthey94}.  As a consequence, continuous effort is done to
find new tools capable of disentangling the age/metallicity
degeneracy, and to give more detailed information on chemical and
physical evolution of galaxies.
On the other hand, our ability to know the evolution of the Universe
relies also on how well we can determine distances up to
galactic/extragalactic/cosmological scales.  In this paper we address
these related topics (the problem of determining the distances and the
physical/chemical structure of the stellar component in galaxies)
using both new observations and new theoretical model predictions of
galaxy integrated colors and {surface brightness fluctuations} (SBF)
properties.

The SBF method was introduced by \citet{ts88} as way of estimating
elliptical galaxy distances.  It took roughly a decade for the most
commonly applied $I$-band implementation of the method to reach
maturity, including a proper calibration in terms of stellar
population \citep[e.g.,][]{tonry97,blake99}. Here, we simply remind
the reader that the method measures the ratio of the second to the
first moment of the stellar luminosity function by analyzing the
spatial fluctuations in the surface brightness of a target galaxy.
This ratio corresponds to the luminosity-weighted mean luminosity of
the stars in the population, which for old stellar populations is
approximately the luminosity of a typical red giant branch star.

Given its definition, it is evident that SBF carries information on
the stellar content of galaxies, similar to classical integrated
spectral and photometric data. However, the information is somewhat
different that given by these classical tools, because of its high
sensitivity to the properties of bright post-{\it Main Sequence}
stars.  Being so, the technique has been widely used in the last
decade, not only to determine distances \citep{tonry01, jensen01}, but
also to study the stellar population properties of galaxies and
globular clusters (\citealt{at94}; \citealt{liu02};
\citealt{jensen03};
\citealt[][Paper I hereafter]{cantiello03}; \citealt{gonzalez04}; 
\citealt[][Paper II hereafter]{raimondo05}).

In this paper we present SBF and broadband color measurements for a
sample of 8 galaxies imaged with the Advanced Camera for Surveys (ACS)
on board of Hubble Space Telescope (HST).  The large format, high
resolution, sharp point spread function (PSF), and good sampling of
the ACS allows for much more detailed studies of galaxy photometric
and SBF gradients than was previously possible.
The paper is organized as follows: Section~2 describes the selected
sample of galaxies and the preliminary data reduction and analysis.
The procedures adopted to derive both surface photometry and SBF
magnitudes are presented in Section~3.  In Section~4, we present our
results and discuss the two aspects of SBF: firstly as distance
indicator, through the calibration of the SBF-color relation; secondly
as a stellar population tracer, with the comparison of observed and
predicted SBF amplitudes and gradients. A summary of the work is given
in Section~5.

\section{Data}

The raw data used for this work are Advanced Camera for Surveys
\citep{ford98,ford03} F435W, F606W, and F814W exposures drawn from HST
archive. All data are taken with ACS in its Wide Field Channel (WFC)
mode.

Most of the observations are deep exposures associated with the
proposal ID \#9427, which was designed to investigate the Globular
Cluster System for a sample of 13 giant ellipticals in the redshift
regime $2000\leq cz \leq 5000$. Among these galaxies we selected six
objects less polluted by dust patches, and with exposure time long
enough to allow SBF measurements.  In the following we will refer to
these galaxies as the ``core sample''.  The data of the other two
galaxies of the sample -- NGC\,404 and NGC\,1344 -- come from programs
\#9293 and \#9399, respectively.

For the core sample, F814W and F435W band exposures are available. For
NGC\,404 and NGC\,1344, we have only F814W, and F814W+F606W exposures,
respectively. Table \ref{tab1} presents a summary of the main
properties for the sampled galaxies: (1) the galaxy name; (2) group
name and group number as defined by \citet{faber89}; (3) right
ascension (J2000) from the RC3; (4) declination (J2000); (5) recession
velocity in the CMB reference frame; (6) morphological T-type from
RC3; (7) B band extinction from \citet{sfd98};
(8) observational program ID number; (9) total exposure time for F814W
images; (10) filter ID of other images available from the same
proposal; (11) total exposure time for the second filter.

To prepare data for analysis, the image processing -- including
cosmic-ray rejection, alignment, and final image combination -- is
performed with the ``Apsis'' data reduction software \citep{blake03}.
In this stage of the data reduction, an RMS image is also generated
for each scientific frame; the value associated to each pixel of this
image is equal to the total pixel noise. This frame will be used in
the photometric analysis of the data.

The latest ACS photometric zero points, and extinction ratios are
applied \citep[hereafter S05]{s05} along with a correction
for Galactic absorption from \citeauthor{sfd98}.  No internal
extinction correction for data has been considered here, both because
of the small quantity of dust in normal ellipticals, and because the
dusty patches sometimes present in our galaxies can be easily
recognized in the bluer bands (mainly $B$-band images) and masked out.

To transform the ACS photometry in the standard UBVRI photometric
system, we use the equations from S05.  We checked the consistency of
the (F435W, F606W, F814W)-to-(B, V, I) transformations by comparing
for each galaxy the transformed magnitudes and colors with available
literature measurements taken in the standard photometric system. For
F606W-to-V, and F814W-to-I transformations, the comparison supports
the reliability of S05 equations. As an example, for NGC\,1344 we
derive $(F606W{-}F814W)_0=0.876 \pm 0.016$~mag which, after applying
the S05 transformations, corresponds to $(V{-}I)_0=1.150 \pm
0.025$~mag to be compared to the \citet[T01]{tonry01} value
$(V{-}I)_0=1.135 \pm 0.011$~mag\footnote{For this comparison we used
the same annular region considered by T01. In particular the color of
the galaxy has been estimated in both cases within a circular annulus
centered on the galaxy and with inner (outer) radius $\sim 6''$
($\sim60''$).}.

Similarly the $(F435W{-}F814W)_0$-to-$(B{-}I)_0$ transformed data for
NGC\,1407, NGC\,4696, NGC\,5322 agree with the data from
\citet{goudfrooij94}\footnote{In order to correctly compare our
measurements with those by \citeauthor{goudfrooij94}, first we have changed
their extinction corrections adopting the \citeauthor{sfd98} dereddening
tables, then we restricted both $(B{-}I)_0$ measurements to the same
annulus (where possible) comparing data obtained within the radii
reported in Tab. \ref{tab7}. In all cases results agree within
$1~\sigma$ level.}.
 
For the range of colors inspected here ($1.81\leq(B{-}I)_0\leq2.25$),
applying S05 equations we derive a correction $(I-F814W) \leq 0.01$~mag
with a typical uncertainty of $\Delta I_{F814W}\sim0.03$, while
for the $B$~band we find on average $(B{-}F435W) \sim 0.03$~mag, and
$\Delta B_{F435W}\sim0.02$.  As a consequence, a systematic
uncertainty of $0.03$~mag in $(B{-}I)_0$ colors comes from the
application of the S05 ACS to standard systems transformations.

Keeping in mind such point, in the following sections all F814W,
F606W, and F814W data will be transformed into standard I, V and B if
not stated otherwise.

\section{Data Analysis}

The SBF data analysis is done following the standard procedure
(\citealt[TAL90 hereafter]{tonry90}; \citealt{jensen98, blake99, mei01}).
Part of the software used for this work, is the same developed for the
SBF distance survey \citep{tonry97}. The main differences are in the
photometry tool adopted -- the software SExtractor \citep{bertin96} --
and in the galaxy isophote fitting procedure, obtained with the
IRAF/STSDAS package ISOPHOTE. In the following sections we will
discuss in detail the stellar and surface photometry procedures, and
the SBF analysis.

\subsection{Photometry}

\textit{(a) Surface Photometry and galaxy subtraction.}

\begin{figure}
\begin{center}
\epsscale{1.}
\plotone{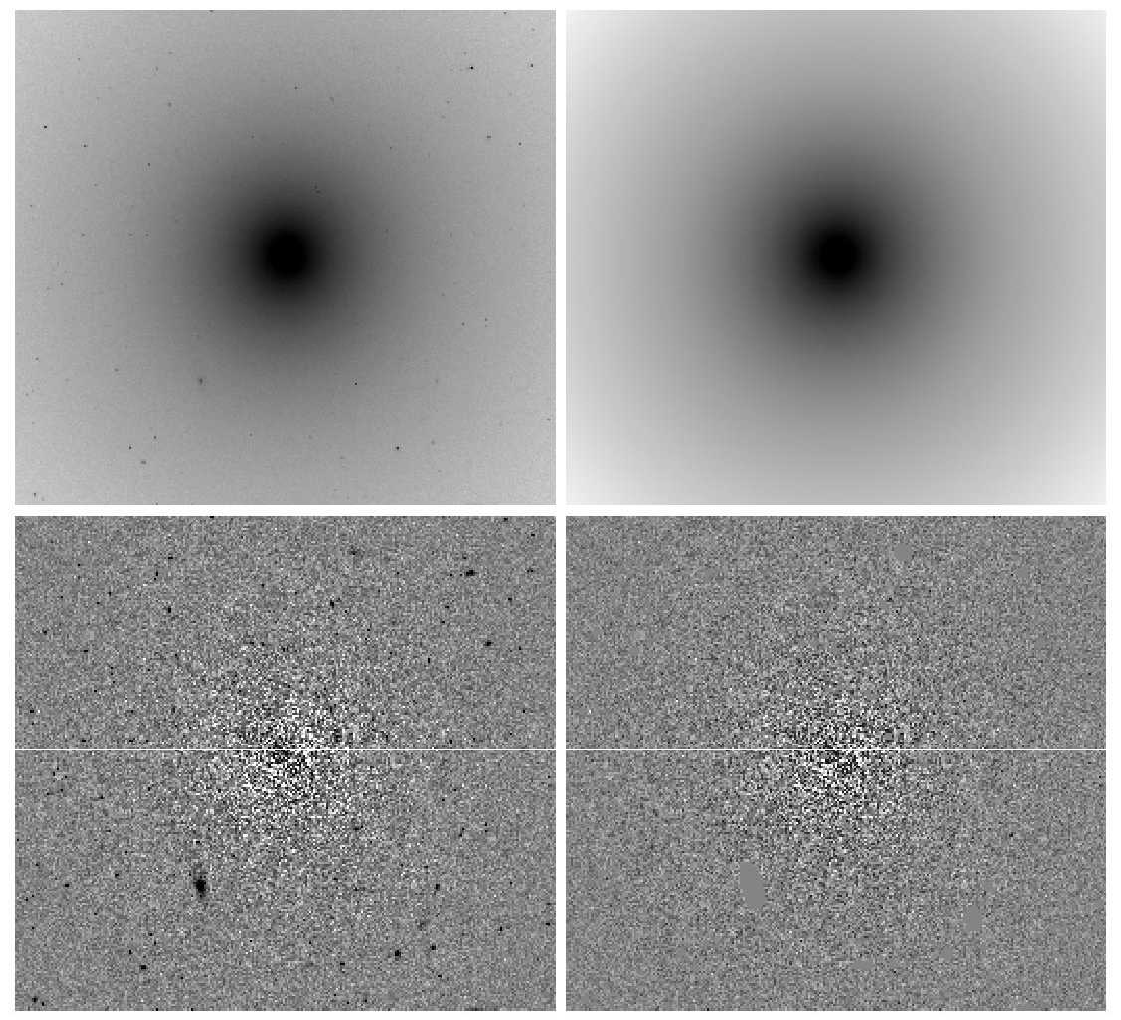}
\vspace{1cm}
\figcaption{The sequence of frames used
for the photometry/SBF analysis (the case of NGC\,1407 is shown).
The upper two panels show the original image (left) and the
galaxy model (right), using a logarithmic stretch.  The
lower panels show the galaxy-subtracted frame (left) and the
final residual frame (right), shown using a simple linear gray
scale.
The rich globular cluster populations can be easily seen after the
galaxy model is subtracted (lower left panel).  In the lower right panel the
external sources visible in the galaxy-subtracted image are masked out, the
fluctuations are clearly visible in the ``bumpiness'' near the center
of this panel.
\label{gxyimg}}
\end{center}
\end{figure}

We choose an iterative process to determine (i) the sky value, (ii) the
best model of the galaxy, and (iii) the external source photometry and
mask. The various steps of the analysis described below are depicted
Fig. \ref{gxyimg}.

The sky contribution in all the galaxies of our sample could not be
simply determined by inspecting the outer part of the CCD, as the
field of view of the detector is not large enough to reach a region of
the sky not contaminated by the light of the observed galaxy.  A
provisional sky value for the whole image is assumed to be equal to
the median pixel value in the corner with the lowest number of counts.
After this sky value has been subtracted from the original image, we
mask out all the obvious sources whose presence can badly affect the
ellipse fitting process, i.e. bright/saturated stars, other galaxies
and, in some cases, dusty regions. The gap region between the two ACS
detectors and other detector artifacts are also masked out.

Afterwards, the fit of the isophotes is performed using the
IRAF/STSDAS task ELLIPSE. The task, which is based on an algorithm by
\citet{jedrzejewski87}, reads an image section and gives as output the
fitted isophotal parameters: semi-major axis length, mean isophote
intensity, ellipticity, position angle, and several others
geometric/photometric parameters.

At this stage we do not set any constraint to the fitting procedure,
all parameters are left free but the initial galaxy center and the
preliminary mask. Besides, the task gives as output the deviation of
the isophotes from the elliptical shape, through the amplitudes of 3rd
and 4th harmonics of a Fourier expansion. Such deviations are small in
all galaxies of our sample except, in some cases, for the very
innermost regions ($< 1$ arcsec) which are excluded from SBF analysis
due either to an inability to fit the steep central intensity profile,
the presence of dust, or both.

Once the preliminary galaxy model has been subtracted from the
sky-subtracted frame, a wealth of faint sources appears
(Fig. \ref{gxyimg}, lower left panel). In the following we will refer
to all these sources -- foreground stars, background galaxies, and the
galaxy's globular clusters -- as {\it external sources}.

All our frames are crowded with these faint objects; in the case of
the Antlia galaxies (NGC\,3258, and NGC\,3268) also a substantial
number of foreground stars are in the frame. To have a reliable
measurement of the galaxy color profile, all unwanted contaminating
sources have to be identified and masked out.

The mask is obtained applying a sigma detection method to the
sky+galaxy subtracted frame. Such detection is done through the
determination of a smooth, spatially variable mean background and
background $rms$, convolving the data with a gaussian kernel to
optimize the detection of the faint sources. Finally this new mask is
combined with the previous one.

The new mask is fed to ELLIPSE to refit the galaxy's isophotes. In
general, if the brightest objects were properly eliminated in the
preliminary mask, this second galaxy model does not differ much from
the previous one, as ELLIPSE itself performs a $\kappa$-$\sigma$
clipping to exclude discrepant points.

\begin{figure}
\epsscale{1.}
\plotone{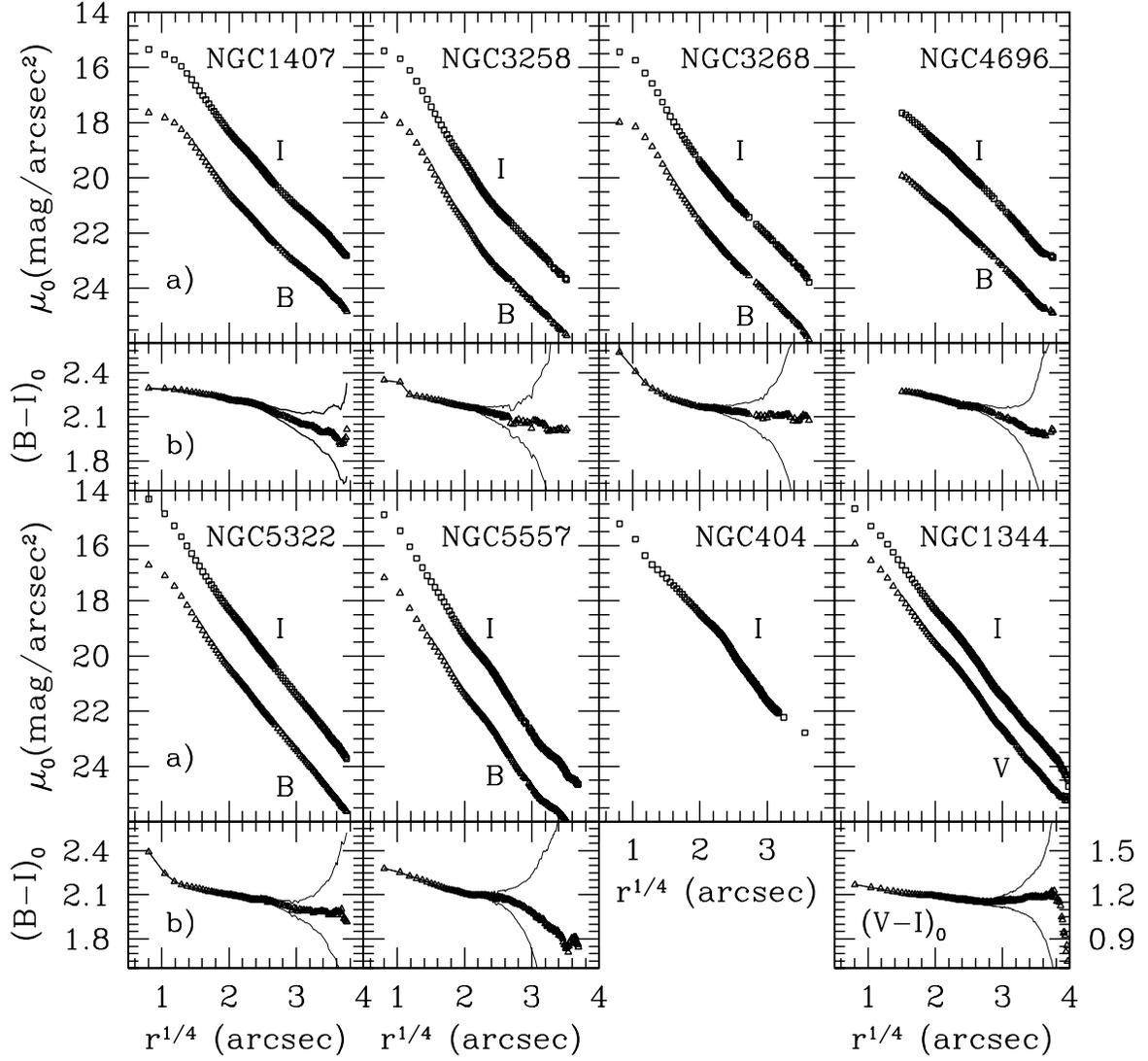}
\figcaption{{\it Panels (a)} - Open squares: I band surface brightness
profile. Open triangles: the same for $B$ band ($V$ band for NGC\,1344).
{\it Panels (b)} - Measured $(B{-}I)_0$ profile as a function of
radius. Error bars are also shown as thin solid lines. Data are
corrected for Galactic extinction, and plotted against the (fourth root of
the) semi-major axis of the isophote. Note that for NGC\,1344 the
$(V{-}I)_0$ color profile is plotted.
\label{sbprofiles}}
\end{figure}

\begin{figure}[t]
\epsscale{0.75}
\plotone{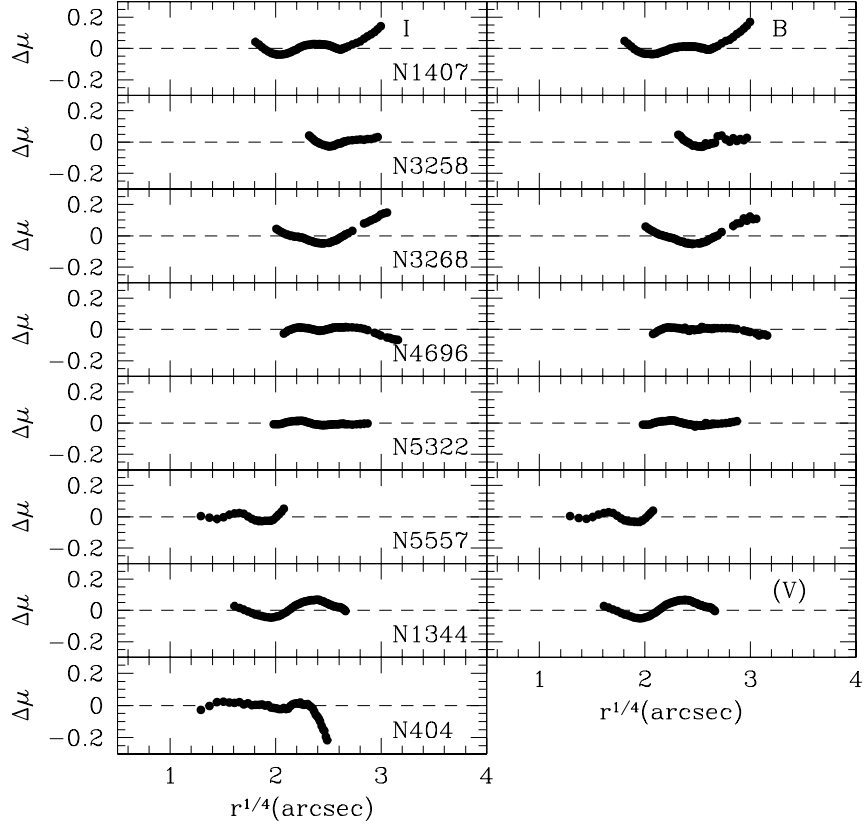}
\figcaption{Difference of the measured
Surface Brightness profile and the $r^{1/4}$ profile fitted in the
outer regions of the galaxies ($\Delta \mu = \mu_{observed} -
\mu_{r^{1/4}fit}$), plotted against galaxy radius. Left (right) panels: I (B) band
profile differences.
\label{deltamu}}
\end{figure}

After having determined the shape of the isophotes, we measure the
flux profile of the galaxy within the modeled isophotal radii. Then,
in order to improve the estimation of the sky level, we fit the
surface brightness profile of the galaxy with a de Vaucouleurs
$r^{1/4}$ law profile plus the constant sky offset. Although the
surface brightness profile of the core sample galaxies is not
$r^{1/4}$-like over the whole range of radii inspected
(Fig. \ref{sbprofiles}, panels {\it a}), we find that for the
outermost modeled regions a de Vaucouleurs' law is well suited
(Fig. \ref{deltamu}; see also \citealt{tonry97}, \S\,1.2).  Thus, to
estimate the image sky background we restrict the fit to these radii;
this value is then adopted as final sky value.  The correction to the
preliminary sky is generally small, less than 10\% of the provisional
value, for all galaxies.

The above procedure is used to determine the sky in all images. For
the purpose of the sky determination only, the geometry of the $I$
isophotes (position angle, center, and ellipticity) is used also when
analyzing the $B$- and $V$-band images. This choice is justified by
the fact that in the $I$-band frame the Galaxy/Sky counts ratio is
higher at the same radius compared to other bands' frames,
consequently the shape of isophotes is better defined, allowing for
lower uncertainty in the sky value estimation. In fact, assuming the
$I$-band shape for the isophotes it is possible to analyze the surface
brightness profile of the galaxy to larger radii also in the $B$-,
$V$-band frame, so that a better power-law fit can be performed. On the
contrary, we fitted and subtracted independently the galaxy in all
frames for the aim of detecting/masking external sources, if
needed. As a matter of fact, while the sources detection is more
efficient in the $I$-band frame, the presence of dust is more easily
recognizable in the B (V) frame, which - after the galaxy subtraction
- is used to define the mask of dusty regions if any.

In panels (a)-(b) of Figure \ref{sbprofiles}, we show the surface
brightness and color profiles for each galaxy.
The profiles do not need any seeing correction due to the high
resolution and narrow Point Spread Function (PSF) of ACS/WFC data,
whose FWHM is 0.10''.

The next step is the subtraction of the galaxy model from the
frame. Even though the model fitting has been subject to an iterative
procedure, the model subtracted frame shows large-scale deviations
from a flat background.  To subtract these large scale residuals, we
use the background map derived running the photometry package SExtractor
(via a combination of $\kappa$-$\sigma$ clipping, and mode
estimation).  For this issue we tune the SExtractor parameters
BACK\_FILTERSIZE and BACK\_SIZE so that the background subtracted
image does not show any large-scale deviation. As well known
\citep{blake99}, this procedure corrupts the lowest wave
numbers of the image power spectrum, which must be omitted in the
following determination of SBF amplitudes.

\textit{(b) Point sources and background galaxies photometry}

Once the sky, the galaxy model, and the large-scale deviations are
subtracted from the original image, we derive the photometry of the
external sources in the frame. The only masked sources at this point
are the saturated foreground stars, and, eventually, other very
extended galaxies present in the frame, if any.

The construction of the photometric catalog is critical for the
estimation of the luminosity function.  As will be shown in the
following section, our aim is to fit the luminosity function of
external sources, so that we can infer (by extrapolation) its faint
end, which is fundamental in order to have reliable SBF
estimations. This is due to the fact that the fluctuations that we can
measure from the (sky+galaxy+large scale residuals) subtracted frame
include also a contribution arising from the undetected external
sources left in the image.

Given the high spatial resolution of ACS and the low background of HST
measurements, the identification of Globular Clusters (GCs), faint
background galaxies, and dust is easier compared to analogous ground
based measurements. The limiting magnitude of undetected point sources
\citep[e.g.,][]{jacoby92} is fainter with ACS data than with
typical ground-based data, consequently the external sources
contribution to SBF is very small.  Nonetheless, the radial gradient
of fluctuation power spectrum that we are looking for, could be also
relatively small, so we need a reliable estimation of such residual
variance, whose evaluation depends on the goodness of the fitted
luminosity function.

To obtain the photometry of external sources we used SExtractor, as it
gives a good photometry of both extended and non extended sources.  It
also gives as output the background map which, as explained before, is used
to remove the large scale residuals of the galaxy subtracted frame,
and it accepts user supplied weight images. The latter point is of
great interest for these measurements, as, by specifying the error map,
the photometric uncertainty can be estimated taking
into account the contribution due to the subtracted galaxy.
We have modified the SExtractor input weight image by adding
the galaxy model (times a factor $\sim 1$) to the RMS image, so that
the surface brightness fluctuations are recognized as noise
(TAL90).  

The main drawback with SExtractor is that its output is as good as the
choice of the input parameters. Although there are more than 30 of
them, the critical ones for us are those related to the
detection, the background estimation, and to the deblending
procedures. In general, the parameters are chosen to minimize at the
same time the number of spurious detections, and the number of obvious
real objects missed by the detection algorithm.  As an example in
Tab. \ref{tab2} we report our SExtractor input parameter file, {\it
default.sex}, used for NGC\,1407.

Among the five different types of magnitudes SExtractor can evaluate
for each source, we consider only two: MAG\_AUTO, for extended
sources, and MAG\_APER, for point-like sources. The first one is the
aperture magnitude, which is mainly intended to give the most precise
estimation of the total magnitude of extended objects, through a
Kron-like method. The second one gives an estimation of the total
magnitude within a circular aperture whose radius is user supplied,
thus it is the proper magnitude to be associated with stellar objects.

The distinction between the two classes of objects (star or galaxy) is
made through the inspection of the FWHM (SExtractor FWHM\_IMAGE flag),
the object semi-minor to semi-major axis ratio (A\_IMAGE/B\_IMAGE),
and the SExtractor star-galaxy classification parameter (CLASS\_STAR).
These parameters are not fully independent; however, a detailed 
inspection of all of these indicators allow us to have a better
star/galaxy identification, and major problems in the classification
only appear with the faintest objects detected in the frame \citep{cantiello04}. 

As a final step in building the photometric catalog, we need to evaluate
the aperture correction (a.c.). For stellar objects, we make the
standard growth-curve analysis \citep{stetson90}; that is, we choose a
number  of the most isolated point sources
in each frame, averaging the magnitude difference obtained
within circular apertures of diameter $d=6$~pix
($mag_{APER}^{d=6pixel}$) and $d=10$~pix
($mag_{APER}^{d=10pixel}$), respectively.  Further, according to the
S05 prescriptions, we add an extra term of correction from
$d>10~pixels$ to ``infinite'' radius ($a.c._{S05}^{d>10~pixel}$). Finally:

\begin{equation}
a.c._{\,tot} =\langle mag_{APER}^{d=6~pixel} -
mag_{APER}^{d=10~pixel}\rangle + a.c._{S05}^{d>10~pixel}.
\end{equation}

For extended faint sources, it is known that SExtractor can lose up to
half of the total light \citep[and references therein]{benitez04},
thus we apply a rough aperture correction using an interpolation
formula from the results obtained by \citeauthor{benitez04} (see their Fig. 9).

Once the aperture correction is applied to the photometric catalog of the
sources in the frame, the next step is to derive the fit of the
luminosity function. As usual, we assume the total luminosity function
to be the sum of a Gaussian shaped Globular Clusters Luminosity
Function \citep[GCLF,][]{harris91}:
\begin{equation}
n_{GC}(m) =\frac{N_{0}^{GC}} {\sqrt{2 \pi \sigma^2}}~~e^{- \frac {(m - m_I^{GC})^2}{2 \sigma ^2}}
\end{equation}
and a power-law luminosity function \citep{tyson88} for the background
galaxies:
\begin{equation}
n_{gxy}(m) =N_{0}^{gxy} 10 ^ {- \gamma m}
\end{equation}
where  $N_0^{GC}$ ($N_{0}^{gxy}$) is the globular cluster (galaxy)
surface density, and $m_I^{GC}$ is the turnover magnitude of the GCLF
at the galaxy distance.

In expression (3) we use $\gamma = 0.34 \pm 0.01$, according to
\citet{bernstein02}, which is consistent with the value of $\gamma
= 0.33 \pm 0.01$ by \citeauthor{benitez04}\footnote{ The real
uncertainty of $\gamma$ is higher than 0.01, as the intrinsic scatter
in the number of galaxies due to field to field variations is
realistically around 10\%.}. For the GC component eq.~(2) we
assume the turn over magnitude of the GCLF to be $M_I^{GC} =-8.5$~mag, and
the width $\sigma=1.4$ \citep{harris01}.

To fit the Luminosity Function (LF) we used the software developed for
the SBF distance survey; we refer the reader to TAL90 and
\citet{jacoby92} for a detailed description of the procedure. Briefly:
a Distance Modulus (DM) for the galaxy is assumed in order to derive a
first estimation of $m_I^{GC}=DM+M_I^{GC}$, then an iterative fitting
process is started with the number density of galaxies and GC, and the
galaxy distance allowed to vary until the best values of $N_0^{GC}$,
$N_{0}^{gxy}$ and $m_I^{GC}$ are found via a maximum likelihood
method.

\begin{figure}
\epsscale{1.}
\plotone{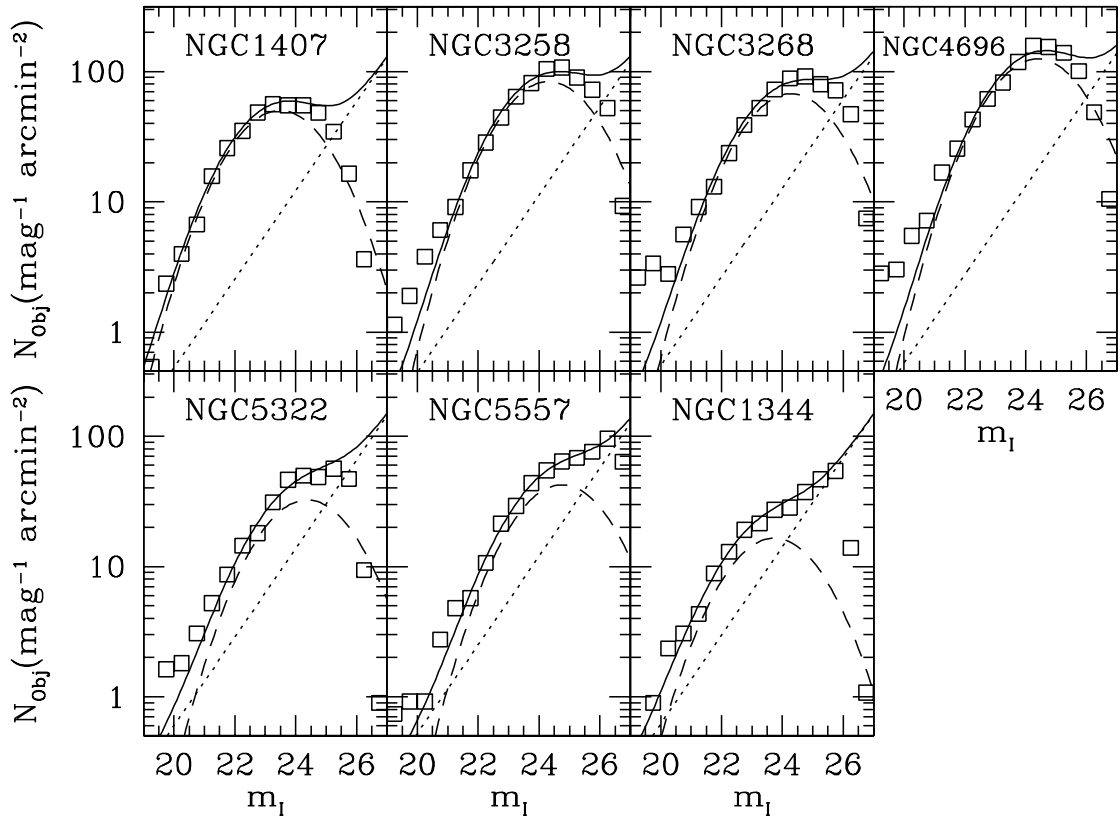}
\figcaption{The fitted Luminosity Function of external sources.  Open
squares mark observational data, the solid line represent the best fit
of the data. The LF of background galaxies and globular clusters are
plotted as a dotted and a dashed line, respectively.
\label{lumfunc}}
\end{figure}

Figure \ref{lumfunc} exhibits the best fit LF of the galaxies in our
sample. As can be noted no LF is derived for NGC\,404, the nearest
galaxy of this sample; in particular no GC (which should be resolved,
spanning few tens of pixels) has been found for this galaxy.

In the error budget section, we will discuss the sensitivity of our
measurements to the details of the assumptions made to build up the luminosity
function. 

\subsection{Surface Brightness Fluctuations}

For a detailed explanation of the SBF method, we remind the reader to
the papers quoted at the beginning of this section. To those we add
the recent papers by \citet{mei05iv, mei05v} on SBF detection
for the ACS Virgo Cluster galaxy Survey, whose procedure is very
similar to the one adopted in this work.

This section is intended to give some hints on the procedure.  In the
following we refer to the (sky+galaxy+large scale residuals)
subtracted frame, divided by the square root of the galaxy model as
{\it residual} frame.

{\it (a) Power Spectrum determination}

The pixel-to-pixel variance in the residual image has several
contributors: (i) the poissonian fluctuation of the stellar counts
(the signal we are interested in), (ii) the galaxy's GC system, (iii)
the background galaxies, and iv) the photon and read out noise.

To analyze all such fluctuations left in the residual frame, it is
useful to study the image power spectrum as all of them are convolved
with the instrumental PSF\footnote{ Since signals convolved in the
real space are multiplied in the Fourier domain, and vice versa.}, but
the noise. We performed the Fourier analysis of the data with the
IRAF/STSDAS package FOURIER.

\begin{figure}[t]
\epsscale{0.95}
\plotone{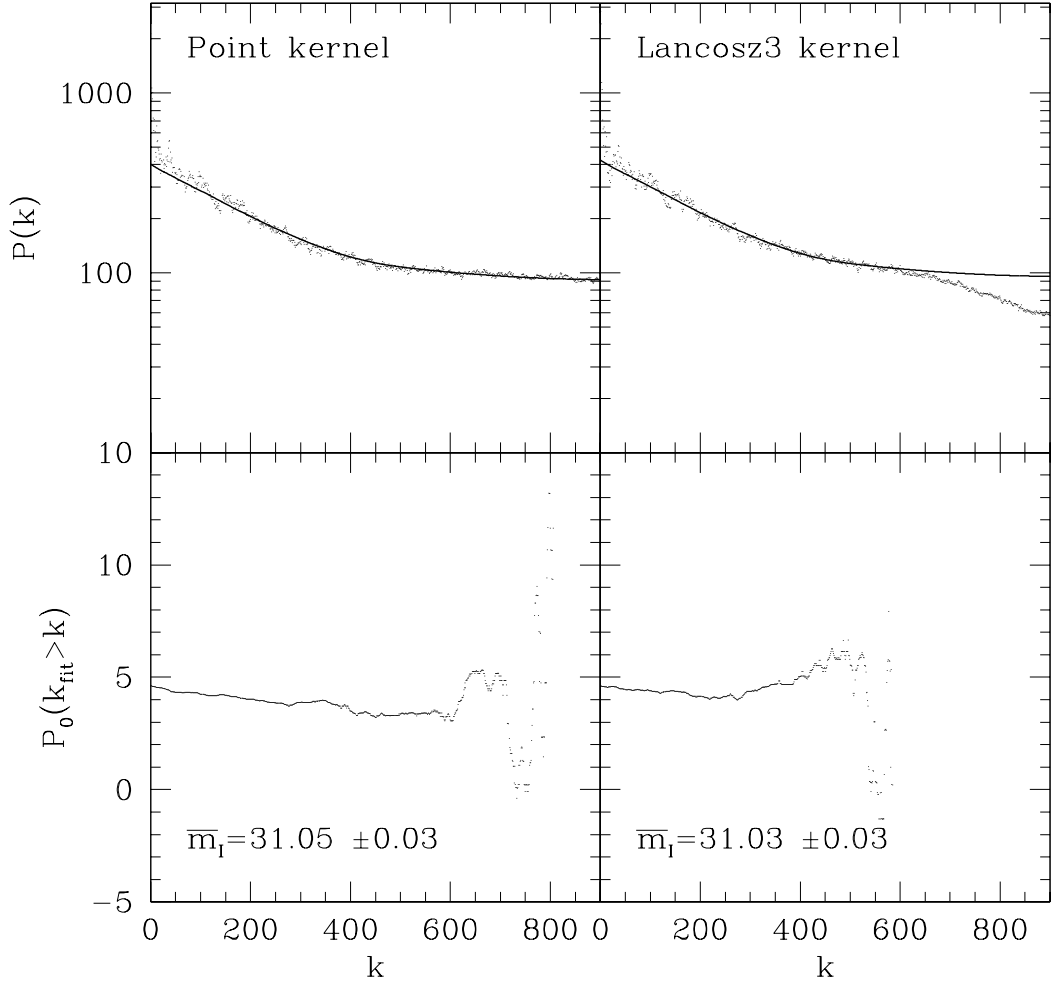}
\figcaption{NGC\,1407 power spectrum analysis. In the upper panels we plot the
azimuthal average of the residual image power spectrum (dots), and the
fit obtained according to the procedure described in text (solid
line). The original images were processed in the same way, but with
different drizzle kernels: Point kernel (left panels), and Lancosz3
kernel (right panels). The lower panels show the fitted $P_0$ as a
function of the starting wavenumber of the fit, $P_0 (k_{fit} > k)$.
The weighted average of the values in the flatter $P_0$ vs. $k$ region
($100\lsim k \lsim 300$), is taken to derive the final $P_0$ value and
its uncertainty. This last data is then adopted to evaluate the final
$\bar{m}_I$, quoted in figure.  All plots refer to the same image
annulus ($20'' \lsim r \lsim 25''$). 
\label{ps}}
\end{figure}

In the Fourier domain the photon and read out noise are characterized
by a white, i.e. constant, power spectrum, thus their
contribution to the fluctuations can be easily recognized as the
constant level at high wave numbers in the image power spectrum (see
Fig. \ref{ps}, upper panels).  On the other hand, since the stellar,
globular clusters, and background galaxy fluctuation signals are all
convolved with the PSF in the spatial domain, they multiply
the PSF power spectrum in the Fourier domain. Thus, the
total fluctuation amplitude can be determined as the factor to be
multiplied to the PSF power spectrum to match the power spectrum 
of the residual frame.

In order to determine this multiplicative constant, we need a well
sampled PSF as reference. Since neither contemporary observations of
isolated stars, nor good PSF candidates are available in our frames,
we used a template PSF from the ACS IDT, constructed from bright
standard star observations.

For the correct estimation of the fluctuation amplitude one must take
into account that the residual frame has been multiplied by a mask
(detected external sources, bad pixels, etc.); this affects the power
spectrum, making it smoother. Thus, the correct reference power
spectrum to be fitted, $E(k)$, results from the convolution of the PSF
and the mask power spectra\footnote{The presence of sharp edges
between masked and unmasked regions of the residual frame could
potentially introduce strong oscillations (ringing) in the image power
spectrum. However, as long as the mean of the image is zero, so that
there is not really a sharp edge, then the ringing is negligible. In
all cases, the power spectra inspected in this paper do not show
detectable presence of such ringing.}.

The fitting procedure is performed in one dimension, thus the
azimuthal average of both the residual frame power spectrum $P(k)$, and the
expectation power spectrum $E(k)$, have to be determined.  The constant that matches the
total fluctuations power spectrum $P(k)$ to $E(k)$ is then evaluated by fitting the
expression:
\begin{equation}
P(k)= P_0 \cdot E (k) + P_1 \,,
\label{eqpk}
\end{equation}
where $P_1$ is the constant white noise contribution, and $P_0$ is the
PSF multiplicative factor that we are looking for.  To compute the
best values for $P_0$ and $P_1$, we use a robust minimization method
\citep{press92}.

{\it (b) Notes on the interval of k used for the fit}

As mentioned before (\S\,3.1.a), the lowest $k$-numbers have to be
excluded when fitting eq.~(\ref{eqpk}), because at these wave numbers
the power spectrum of the residual frame has been corrupted by the subtraction
of the smooth background profile, i.e. by the large scale correlation
introduced by the smoothing.

In addition, for ACS data the very high $k$ numbers must be rejected
too.  In fact the drizzling procedure used to correct ACS images from the
geometric distortion, introduces a correlation in the noise of
adjacent pixels \citep{fruchter02}.

To analyze how this affects the final SBF values, we have made a test
processing the images of several galaxies of our sample adopting two
different drizzle kernels: the reference Lancosz3, and Point kernel
(other commonly used kernels, such as linear or Gaussian, create
severe noise correlations, making them useless for SBF; for further
discussion see \citealt{mei05iv}). As shown in Fig. \ref{ps} (left
panels), the Point kernel does not introduce any correlation in the
noise of adjacent pixels, as demonstrated by the flat power spectrum
at high wave numbers. On the contrary the Lancosz3 drizzle kernel --
which introduces a sinc-like interpolation pattern between the pixels
-- shows a substantial displacement from the expected white noise
power spectrum at high $k$s (Fig. \ref{ps}, right panels).

We performed the fit of eq.~(\ref{eqpk}) using different $k$ intervals of
for each drizzle kernel. For the Point kernel data fit we used
all the $k\geq 100$ wave numbers data, while for the Lancosz3 kernel we also
restricted to the $k$-numbers where no small-scale correlation appears
($k\leq 600$).  Even so, the differences in the fitted
$P_0$ are negligible, and the final $\bar{m}_I$ values for both kernels
agree within the associated uncertainty
(Fig. \ref{ps}).  

We have finally adopted the Lancosz3 as reference drizzle kernel,
since the Point kernel leaves in the processed
frame a large number of null pixels, which adversely affects the
photometry of point sources, and it causes problems for the galaxy
modeling.

{\it (c) External source correction and final $\bar{m}_I$ measurements}

The fluctuation amplitude $P_0$ estimated so far contains the extra
contribution of unmasked external sources.
To reduce the effect of this ``spurious'' signal, all the sources
above a defined signal to noise level (typically we adopted a $S/N
\sim3.5$) have been masked out before evaluating the residual image
power spectrum. Thus, at each radius from the galaxy center a well
defined faint cutoff magnitude ($m_{lim}$) fixes the magnitude of the
faintest objects masked in that region.  Such masking operation
greatly reduces the contribution to $P_0$ due to the external
sources, but the undetected faint and the unmasked low S/N
objects could still significantly affect $P_0$, thus their
contribution -- called the residual variance $P_r$ -- must be properly estimated
and subtracted. The residual variance is computed evaluating the
integral of the second moment of the luminosity function in the flux
interval $[0,~f_{lim}]$:

\begin{equation}
\sigma_r^2 = \int_0^{f_{lim}} N_{Obj}(f) f^2 df
\end{equation}
where $f_{lim}$ is the flux corresponding to $m_{lim}$, 
and $N_{Obj}(f)$ is the luminosity function previously fitted
(\S\,3.1.b).  The residual variance $P_r$ is then $\sigma_r^2$
normalized by the galaxy surface brightness.
For most of the galaxies in our sample, the $P_r/P_0$ ratio
ranges from $0.05$ to $0.15$ in the regions considered for SBF
measurements (it is negligible for NGC\,404).  
A default $25\%$ uncertainty is associated to $P_r$ (TAL90).

Finally, the SBF magnitude is measured from $P_0$, subtracting the
residual contribution evaluated above:

\begin{equation}
\bar{m}_I = -2.5 log (P_0 - P_r) +  m_{zero}^{ACS} + 2.5~log~(Exposure~time)
\end{equation}
where $m_{zero}^{ACS}$ is the zeropoint ACS magnitude reported by S05.

As we are interested also in the study of the SBF radial behavior, the
procedure described above is applied to several (typically 5) {\it
elliptical} annuli in which we divided each galaxy\footnote{ We
compared our SBF measurements, with the results obtained using
circular annuli -- usually adopted for these works. The difference in
the computed fluctuation amplitude increases with the ellipticity but
is less than $0.1$~mag in all cases.  Moreover, the $\bar{m}_I$
vs. $(B{-}I)_0$ profile, which is what we are most interested in, is not
affected by this change.  We finally decided to analyze our data
within elliptical annuli, taking into account the real isophotal
geometry of the observed object in order to study its
photometry.}. The annuli shape reflects the geometry of the isophotes
profile.

Table \ref{tab3} reports the final results of our measurements for
each annulus and for all galaxies of the sample; columns give: (1) the
annulus number, starting from the innermost one; (2) the average
annulus radius; (3) the average annulus ellipticity; (4-5) the
corresponding $(B{-}I)_0$ color; (6-8) the SBF magnitudes before
($\bar{m}_I^{uncorr.}$), and after ($\bar{m}_I$) applying the external
sources correction.

\begin{figure}
\epsscale{1.}  \plotone{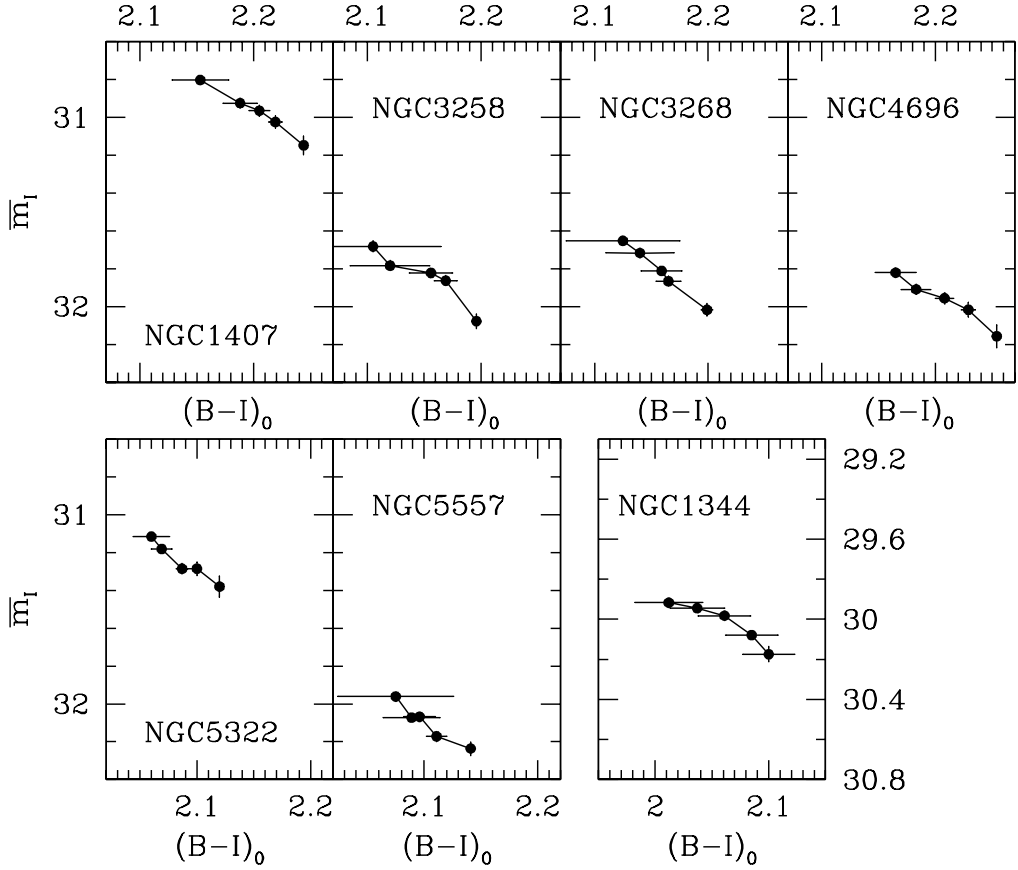} \figcaption{SBF amplitudes measured in
the various galaxy annuli plotted against the $(B{-}I)_0$ color of the
annulus. Note that while the $(B{-}I)_0$ color in
Fig. \ref{sbprofiles} has been evaluated in annuli with semi-major
length: $(r_{out}-r_{in})\sim16$ pixels ($\sim32$ pixels for the
outermost regions), the annuli adopted to evaluate the $(B{-}I)_0$
plotted in this figure are the same used for SBF evaluation, that is
$(r_{in},r_{out})$=[(128, 256), (257, 384), (385, 512), (513, 768),
(769, 1024)] in pixel units.  We advise that in the panels of this figure
(as well as in Fig. 7--11) brighter fluctuations magnitudes,
and bluer $(B{-}I)_0$ colors in general correspond to more external
annuli. For sake of clearness the plotted axis range is different for
each of the three blocks in the figure, according to the observed
ranges of $(B{-}I)_0$ and $\bar{m}_I$ for the galaxies, though we keep
fixed the amplitude of the intervals. Concerning NGC\,1344 see
Appendix A for the color equations adopted to obtain the $(B{-}I)_0$
from the measured $(V{-}I)_0$.
\label{results}}
\end{figure}

All these data are plotted in Figure \ref{results}, which presents the
measured $\bar{m}_I$ vs. $(B{-}I)_0$ profiles. With regards to this
figure - and figures 7--11 - it is worth to note that brighter SBF
magnitudes, and bluer $(B{-}I)_0$ colors correspond to outer galaxy's
regions.

\subsection{Error Budget}

Before concluding this section, we intend to discuss how the results
obtained are affected by substantial changes of some of the assumption
made.  In particular we will discuss how variations in the (i) PSF,
(ii) sky value, and in the (iii) luminosity function affect the
estimated $\bar{m}_I$ and $(B{-}I)_0$.

{\it (i) PSF} - The ACS Point Spread Function shows to be stable
enough in time. It is also fairly stable over the ACS field of view
\citep{krist03}, with small spatial variation that is not significant
for our analysis \citep{mei05iv}.

The PSF used for this work comes from the ACS IDT, and is derived from
the composition of several well isolated point sources obtained from
other ACS observations.  In order to study the sensitivity of the
$P_0$ values to the PSF assumed, we fitted eq.~(\ref{eqpk}) adopting
three different PSFs, derived from individual high signal-to-noise stars
in archival F814W ACS images. As a result we have obtained that the new $P_0$
derived agree within $2-3\%$ with the earlier values. Consequently, the
final estimated $\bar{m}_I$ contains a systematic uncertainty due to
the choice of the PSF, which is of the order of $0.03$~mag.

{\it (ii) Sky } - The estimation of sky background for the whole set
of galaxies images is done under the assumption that the de
Vaucouleurs $r^{1/4}$ profile describes the outer parts of the galaxy
radial profile.  However, such assumption may fail in the case of
bright ellipticals, which are typically best fitted with a Sersic
$r^{1/n}$ profile \citep{caon93}. We have verified how much the
estimated sky would change if we assumed a $r^{1/n}$ profile. As a
result we have that the changes are within $1~\sigma$ for the $n$=[3-8]
range (which is still reasonable for these bright galaxies) but would
go to $2~\sigma$ for $n$=2. Thus, we performed a test by changing the
sky values within $2~\sigma$ the original value, the new values of
$(B{-}I)_0$ and $\bar{m}_I$ measured agree within the associated
uncertainty with the original measurements. Such variations are even
less dramatic if the inner annuli are considered as well. Here, in
fact, the sky to galaxy counts ratio is generally $\leq0.05$, and the
estimated original/perturbed colors and SBF agree under sky variations
as high as $5~\sigma$ the original value.

{\it (iii) Luminosity Function} - To verify how the uncertainty in the
fitted LF affects our results, we have performed several tests. First
we have changed various SExtractor detection parameters, then we have
also changed the convolution kernel for point sources detection (see
\citealt{bertin96} for the details) and the parameters for the
background subtraction.  Finally, we have adopted a different routine
to fit the LF, using the one from the ACS Virgo Cluster Survey which
is optimized for dealing with ACS data. The resulting photometric
catalogs differ from the original ones, but the final $\bar{m}_I$
values are well within $1~\sigma$ of the original measure.  However,
the result of this test is not surprising, since the high quality of
ACS data allows to have an accurate photometric catalog of external
sources down to the faintest magnitudes. Thus, the contamination due
to undetected or low S/N sources is quite low, and residual $P_r$ does
not give a major contribution to the fluctuations (compare the
$\bar{m}_I^{uncorr.}$ and $\bar{m}_I$ values in column 6--7 of
Table 3).

We have also tested how changing the slope $\gamma$ in the magnitude
distribution of the background galaxies affects SBF magnitudes. For
this test we used the $\gamma$ value determined by
\citet{benitez04}. As a result we have found that any detectable
effect exists, the variation being well below $\sim0.01$~mag. However,
it must be emphasized that the $\gamma$ estimations from
\citet{bernstein02}, and \citeauthor{benitez04} are consistent each other,
and differ less than 5\%.

Finally, to complete the list of uncertainties, it is worth noting
that there is a $\sim 0.01$~mag error from the filter zeropoint, and a
$\sim 0.01$~mag error from the flat fielding.  To them we must add the
mentioned uncertainty due to the transformation from ACS filter system
to the standard system, which introduces a systematic error of
$\sim0.02$ ($\sim0.01$, $\sim0.03$) in $I$-band ($V$-, $B$-band)
measurements\footnote{ For some more details on the subject of this
section, see \citet{cantiello04} PhD thesis, available at the web site
http://www.te.astro.it/osservatorio/personale/cantiello/homepage.html}.

\section{Discussion}
In the following sections we will discuss separately our SBF
measurements as a tool to study distances, and to trace the properties
of the dominant stellar populations of galaxies.

\subsection{Calibration and Distances issues}

\begin{figure}[t]
\epsscale{0.75}
\plotone{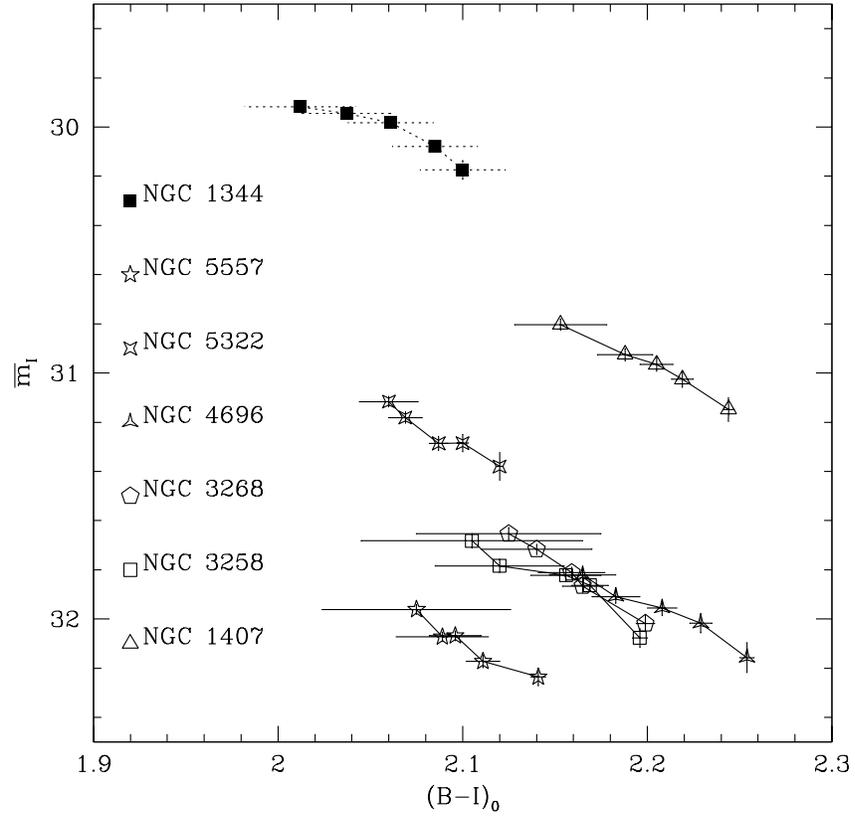}
\figcaption{The distribution of apparent SBF magnitudes versus galaxy
color.  NGC\,404 is not displayed for the clearness of the plot, it is
located at [$(B{-}I)_0\sim 1.81$, $\bar{m}_I\sim-2.2$].
\label{mapparent}}
\end{figure}

The application of the SBF method as distance indicator relies on the
calibration of $\bar{M}$ versus the galaxy broadband color (TAL90;
\citealt{tonry97}). In our case we need to determine the slope
$\beta$, and the zeropoint $\alpha$ of the equation:
\begin{equation}
\bar{M}_I=\alpha + \beta \cdot[(B-I)_0-2.0]
\label{eqzero}
\end{equation}
from the sample of available data\footnote{According to equation A1,
the reference color $(B{-}I)_0=2.0$~mag corresponds to the value of
$(V{-}I)_0\sim1.15$ typically taken as reference color for the
$\bar{m}_I$ vs. $(V{-}I)_0$ calibrations
(e.g., \citealt{tonry97};T01).}. Figure \ref{mapparent} presents
the complete dataset of measurements $\bar{m}_I$ vs. $(B{-}I)_0$
derived according to the procedures described in the preceding
sections\footnote{ To derive the $(B{-}I)_0$ color of NGC\,1344 and
NGC\,404 we need to adopt some $(V{-}I)_0$-to-$(B{-}I)_0$ color
transformation, as we only have $V$- and $I$-band data. To this aim,
we used the relations obtained by adopting an upgraded version of the
\citet{cantiello03} models for simple stellar populations (see
Appendix A).}.

\begin{figure}[t]
\epsscale{0.75}
\plotone{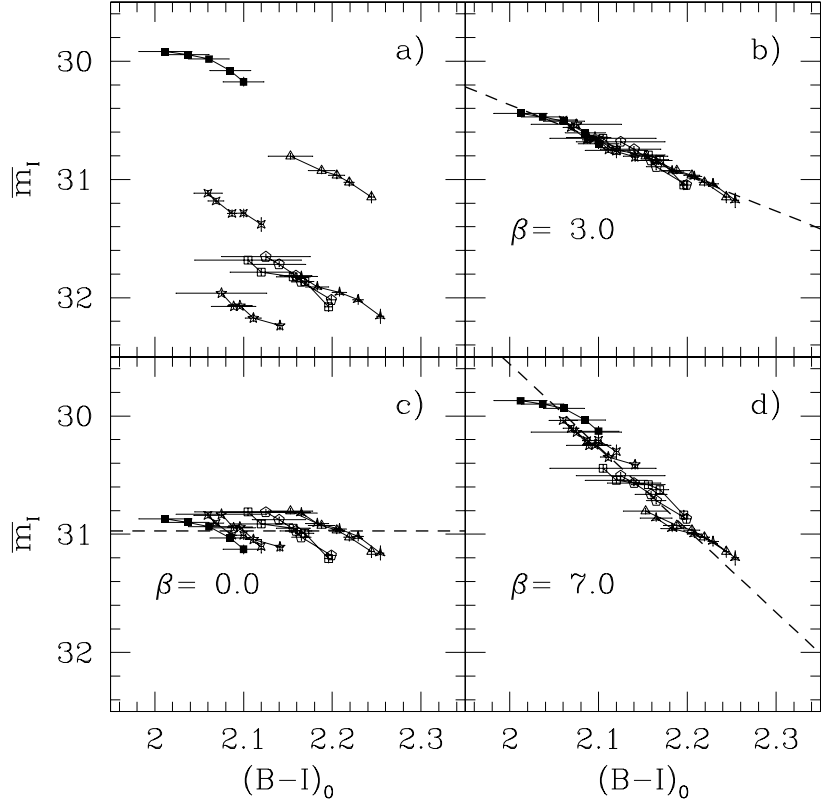}
\figcaption{Panel $(a)$ - The observed distribution of
$\bar{m}_I$ vs. $(B{-}I)_0$ data for the whole sample of galaxies, but
NGC\,404. Panel $(b)$ - Observational data are shifted to the
reference (relative) distance of NGC\,1407. The best fit to the data
is plotted as a dashed line. The value of $\beta$ assumed for the
shifting is reported too. Panels $(c)$, and $(d)$ - similar to panel
$(b)$ but for different $\beta$ values. Our best estimation of the
slope is $\beta\sim3$ (panel $b$).  By comparing the $(b)$, $(c)$, and
$(d)$ panels, it can be seen how the minimization procedure works: the
best fit value of $\beta$ is the one that minimizes the scatter of
observational data with the expected linear dependence.
\label{calibra}}
\end{figure}

In order to determine the best values of $\alpha$, and $\beta$, we
simultaneously fit all the galaxies of the sample, but NGC\,404 for
which we have no independent (radial) color information. The method
applied to fit the unknown parameters of eq.~(\ref{eqzero}) consists of
the following steps. (a) Fix a lower limit value of $\beta$, then
evaluate:
\begin{equation}
\langle\bar{m}_I\rangle_j=\langle \bar{m}_I^{(i)} - \beta \cdot[(B-I)_0^{(i)}-2.0]\rangle_j
\end{equation}
by averaging over all annuli (index $i$) for each galaxy (index $j$).
The value $\langle\bar{m}_I\rangle_j$ corresponds to the mean apparent
fluctuation magnitude of the galaxy, at the fiducial color
$(B{-}I)_0=2.0$~mag.  Note that this quantity is strictly dependent on
$\beta$.  (b) Shift all galaxies to a common value of
$\langle\bar{m}_I\rangle_{Reference}$ according to their
$\langle\bar{m}_I\rangle_j$.  (c) Calculate the $\chi^2$ over all
annuli of all galaxies. (d) Increase $\beta$, and repeat the steps (a),
(b), (c).  The final value of the slope is assumed to be the one that
minimizes the $\chi^2$ value (see \citealt{cantiello04} for full details; a
similar approach is adopted in \citealt{tonry97}).  As an
illustration, Figure~\ref{calibra} exhibits the whole sample of data
before (panel $a$) and after the shifting with three different values
of $\beta$ (panels $b$, $c$, and $d$), keeping the NGC\,1407 distance
fixed.
Adopting this procedure, we derive a best-fit value of $\beta=3.0$,
shown in Figure~\ref{calibra}$b$. To obtain a realistic
estimate of the uncertainty of the slope, we perform a series of tests
by applying a bootstrap method \citep{press92}. In conclusion the
slope $\beta$ and its estimated uncertainty are $3.0 \pm 0.3$.

Apart from determining $\beta$, the procedure applied allows to derive
the relative distance moduli between our sample galaxies, without any
assumption but the best value for $\beta$. In Table \ref{tab4} we
summarize the results obtained, taking NGC\,1407 as reference galaxy.

Once we have evaluated $\beta$ and all relative distance moduli, to
determine the zeropoint $\alpha$ we assume the Fornax cluster DM as
reference -- NGC\,1344 being a member of this cluster.  Adopting
$DM_{Fornax}=31.5 \pm 0.1$ (\citealt{ferrarese00}; T01;
\citealt{blake02}) we obtain: $\alpha=-1.6
\pm0.1$\footnote{ The same result is obtained if the we consider as
reference the Eridanus cluster (NGC\,1407), with a distance modulus
$32.0 \pm 0.1$ (\citealt{ferrarese00}; T01).}.
Finally, we have:
\begin{equation}
\bar{M}_I=(-1.6\pm0.1)+(3.0\pm0.3)\cdot[(B-I)_0-2.0].
\label{eqcal}
\end{equation}

To test this result we compare it with the $I$-band calibration
derived by \citet{ajhar97} for WFPC2 HST data. After applying the
$(B{-}I)_0$-to-$(V{-}I)_0$ transformation equations reported in
Appendix A (eq.~\ref{eqbivi}), eq.~(\ref{eqcal}) becomes:

\begin{equation}
\bar{M}_I=(-1.6 \pm 0.2) + (6.5 \pm 0.6) \cdot[(V-I)_0-1.15],
\label{eqtrans}
\end{equation}
in fair agreement with the derivation by \citeauthor{ajhar97}:
\begin{equation}
\bar{M}_I=(-1.73 \pm 0.07) + (6.5 \pm 0.7) \cdot[(V-I)_0-1.15].
\end{equation}

Having fixed the Fornax Cluster DM as reference, we can also derive
the galaxies' absolute distance moduli from relative DM in Table
\ref{tab4}. The resulting absolute DM are reported in column (3) of
Table \ref{tab4} ($DM_{This\,Work}$)\footnote{For comparison we report
in columns 4-5 the DM obtained from the weighted average of various
other distances estimations (as reported in Table \ref{tab6}, see
\S\,4.2.2 for some details on the table content). It is worth to note
the agreement between SBF-gradient based distances and the other
estimations, both if single galaxy distances are assumed, or group
distances.}.

As a secondary check, we test our empirical calibration deriving the
distance modulus of NGC\,404.  It must be pointed out that such application
need to be considered with some cautions. In fact eq.~(\ref{eqcal}) has
been obtained for galaxies in the color range $2.00\leq(B-I)\leq2.25$,
while NGC\,404 is outside this interval, having $(B{-}I)_0=1.81\pm0.04$~mag.
In spite of that, for this galaxy we derive $(\bar{m}_I-\bar{M}_I)=27.59 \pm
0.15$, in agreement with measurements from the ground based
SBF survey ($27.57\pm0.10$, T01), and with the estimation given by
\citet{tikhonov03}, who obtain
$27.67\pm0.15$ through the TRGB method.

On the theoretical side, adopting the \citet[hereafter
BVA01]{bva01} stellar populations models in the age range
$4-18~Gyr$, for metallicity $-0.4 \leq [Fe/H] \leq 0.2 $, and fitting
a straight line to these data we find $\beta=3.5 \pm 0.2$, and
$\alpha=-1.75\pm0.1$. While, considering their composite populations
models, the value of $\beta$ can reach values down to $2.7\pm0.2$,
depending on the choice on the age and chemical properties of the
stellar population mix (the zeropoint is left practically unchanged).

These results are confirmed by our new SBF models\footnote{The new
models are available at the web site of the Teramo--SPoT group:
http://www.te.astro.it/SPoT} which give $\beta=3.3
\pm0.3$, and $\alpha=-1.7\pm0.1$.  The latter models are derived as
described in Paper I, by using an upgraded set of evolutionary tracks
\citep{pietrinferni04} and a slightly different approach in
considering the evolution of post-AGB stars (Paper II).

\subsection{Stellar Populations issues}

Combining SBF magnitudes and color data with synthetic stellar
population models, one can infer information on the stellar content of
galaxies (e.g., TAL90; \citealt{buzzoni93}; \citealt{brocato98}; \citealt{worthey93};
\citealt{blake99}).  Recent studies in this regard are broadly
consistent with the view that ellipticals are generally not composed
of a coeval, single metallicity stellar population (BVA01; Paper I;
\citealt{jensen03}), supporting results given by spectrophotometric
indicators \citep[e.g.,][]{kobayashi99, trager00i, trager00ii}.

The difficulty of using SBF as stellar population tracers, is the need
of an independent estimation of the galaxy distance, if multi-band SBF
measurements (that is SBF colors) are not available. However, having
measured SBF magnitudes within various annuli of a single galaxy, we
can study the radial behavior of the SBF vs. $(B{-}I)_0$ relation, or
the radial SBF variations within each galaxy, in both cases the
analysis of the relative changes is independent of galaxy distance.
We divide the following discussion into {\it distance independent}
and {\it distance dependent} analyses of the galaxy stellar populations.

\subsubsection{Distance independent analysis: SBF--color relations
and radial gradients}

Radial gradients in broadband colors and line strength indices have
long been used in the effort to constrain galaxy formation models
\citep[e.g.,][]{strom76, cohen79, franx89, davies93}. This is because
different formation histories can produce very different radial
gradients.  For instance, the traditional monolithic collapse scenario
predicts steep inward metallicity gradients, which become more
pronounced for more massive objects, while hierarchical merging models
predict that radial gradients will flatten as galaxies undergo
mergers; thus the most massive galaxies have shallower gradients
because they experience the most merging \citep[e.g.,][]{white80, bekki01, 
labarbera03, labarbera04}. Thus, by characterizing
the radial gradients in multiple stellar population indicators, it is
possible to obtain useful information on a galaxy's formation history.

Here, we consider the information provided on stellar population
variations by our measured SBF and color gradients, and, in
particular, from the slopes of the individual SBF--color relations.
In \S\,4.1 we have derived an ``overall'' slope for the $\bar{M}_I$
vs. $(B{-}I)_0$ relation. The method applied to derive the value of
$\beta$ allowed us to study this relation by using the SBF and color
measurements of the whole sample of galaxies. The drawback in such a
procedure is that it obscures the peculiarities of single
objects. Here we intend to study each galaxy separately, pointing out
the intrinsic differences among them, and suggesting the possible
physical origin of these differences.

As can be recognized by eye inspection of Fig. \ref{mapparent}, the
slope of $\bar{M}_I$ vs. $(B{-}I)_0$, does not show strong variations
on the whole set of plotted data.  However, applying a weighted least
squares method to derive the slope $\beta$ for each galaxy separately,
some differences emerge. Table \ref{tab5} list the slopes estimated,
the reduced $\chi^2$, and the values of $q$ computed\footnote{ The
quantity $q$ is the probability that the $\chi^2$ exceeds a particular
value by chance, i.e. it gives a quantitative measure for the
goodness-of-fit: a $q\sim1$ means that the model reliably fits the
data and it can be accepted, for $q\sim0$ the model must be rejected
\citep{press92}.} for single galaxies.

Two main considerations arise by inspecting the data in Table
\ref{tab5}. First, there is a general agreement between these slopes
within the quoted uncertainties. Moreover, with the exception of
NGC\,3258, all other listed values agree with the overall slope
derived in the previous section. As discussed in more detail below,
the steeper slope of the SBF-color relation in NGC\,3258 would be
appropriate for a pure metallicity gradient at fixed age, while the
significantly shallower slope found for NGC\,1344 requires some
gradient in stellar population age.  This is noteworthy because
NGC\,1344 is the bluest galaxy in our sample with the only exception of
NGC\,404, for which we do not have enough data to measure an SBF
gradient; it also shows morphological irregularities indicative of
recent merging activity \citep{carter82}. Since fractional age
gradients will lessen with time (while metallicity gradients will
remain fixed in the absence of further activity), it is reasonable
that the SBF-color relation would suggest an age gradient in the
youngest, bluest galaxy.
Ultimately, we hope to correlate the slopes of SBF--color relations in
large galaxy samples with other observables, including the local 
environment and  other stellar population indicators,
such as metal and Balmer absorption line strengths,
in order to gain further insight into internal stellar population
variations and enrichment histories.

\subsubsection{Distance dependent analysis: absolute SBF magnitudes}

\begin{figure}[t]
\epsscale{0.75} \plotone{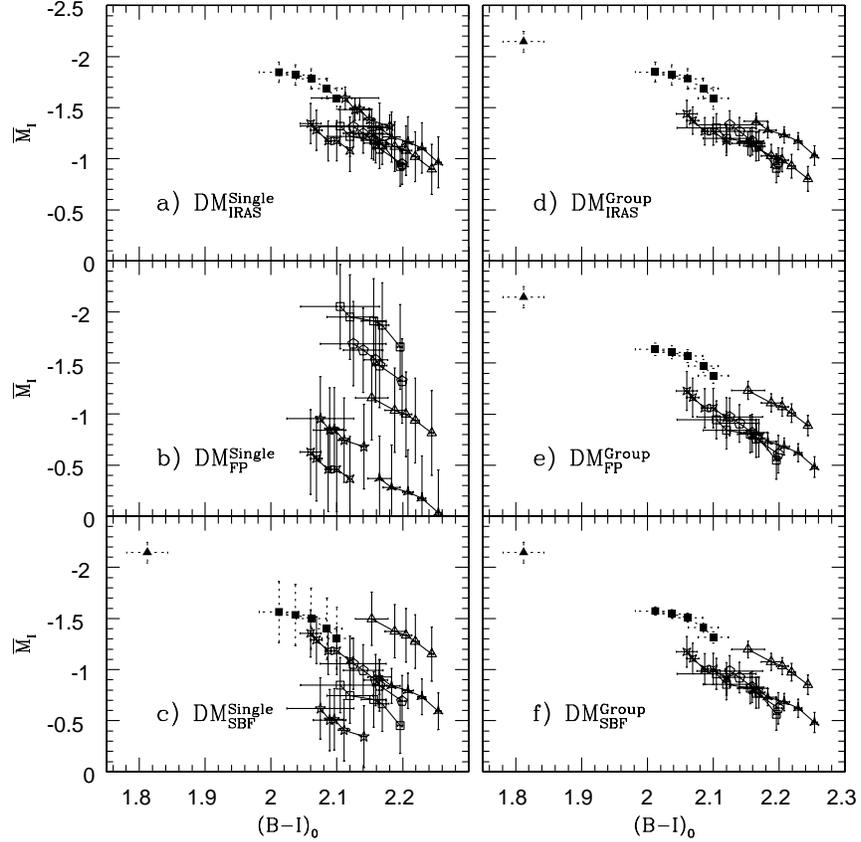} \figcaption{The distribution of the
absolute SBF magnitudes as a function of $(B{-}I)_0$. Each panel
refers to a different assumption on the distance modulus (upper left
quote).  Left panels refer to the distances derived for the single
object, right panels to group average distances (for some useful notes
on group definitions see T01). Symbols are the same of
Fig. \ref{mapparent}.  For NGC\,404 (full triangle at $(B{-}I)_0\sim
1.81$, $\bar{m}_I\sim25.4$) no group distance is assumed, the SBF
distance modulus is adopted in all cases.
\label{mabsolute}}
\end{figure}

To make a direct comparison of the single-band SBF data with predictions
from models, we need to assume distance moduli for the galaxies.
In Table \ref{tab6} we report several distance moduli for each galaxy,
estimated by using three different distance indicators: SBF (ground
based measurements from T01), Fundamental Plane (FP), and distances
derived from the IRAS redshift survey density field (the last two are
from \citealt{blake02}, see the paper for the details). Further,
both single and group distances are considered.  We show in
Fig.~\ref{mabsolute} the location of our galaxy sample in the
$\bar{M}_I$ vs. $(B{-}I)_0$ plane, against the various assumptions on
the distance moduli.
Clearly, the dispersion among the various derivations of distance moduli 
in Fig.~\ref{mabsolute} is quite high, in particular if single FP and SBF
galaxy distances are used (panels $(b), (c)$).

\begin{figure}[t]
\epsscale{1.0}
\plottwo{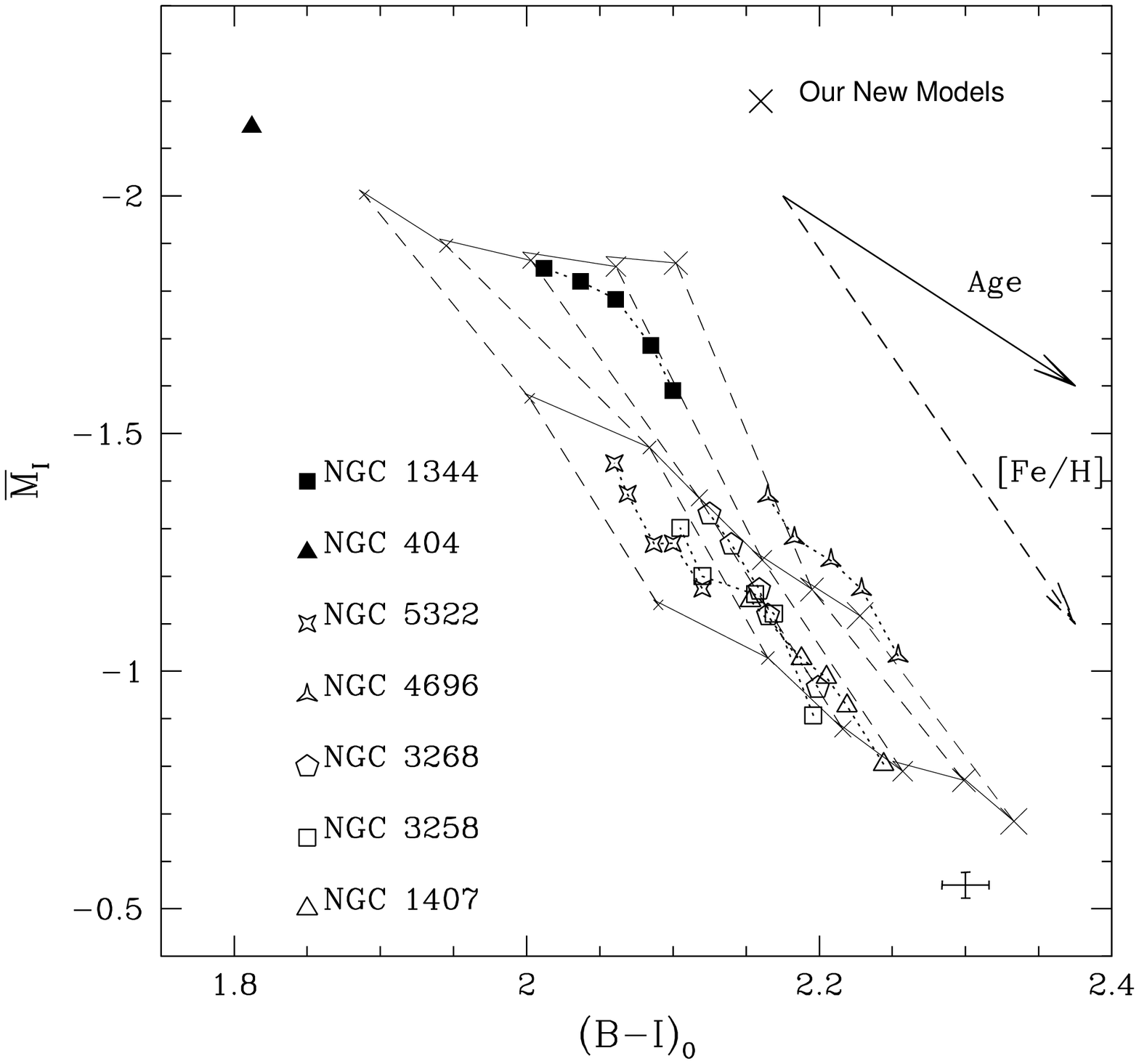}{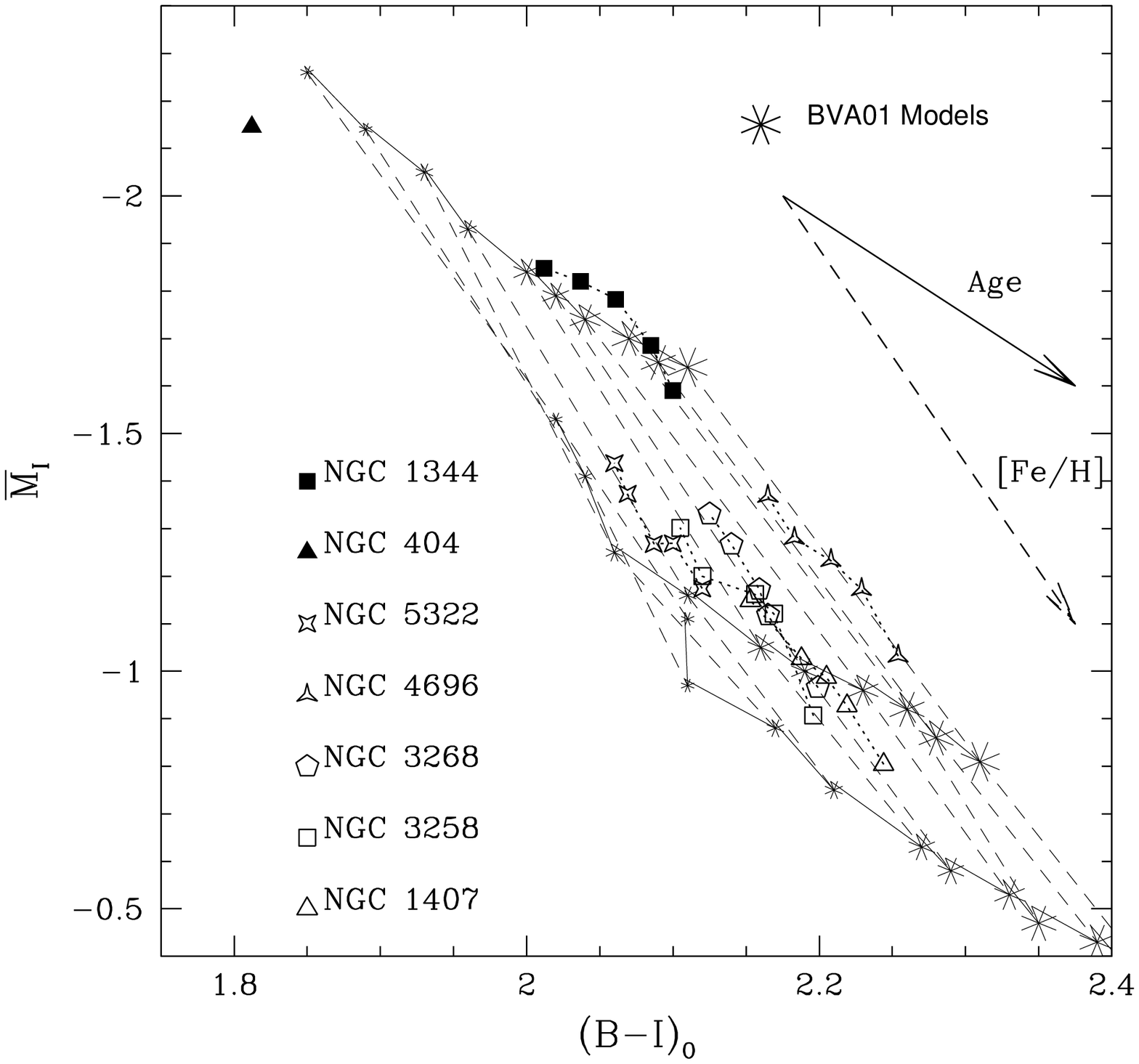}
\figcaption{Absolute $\bar{M}_I$ magnitude derived assuming the IRAS
group distances (in all cases but NGC\,404; see text for details)
versus the measured $(B{-}I)_0$ of the sample. {\it Left panel} - Our
new SBF models, for [5,7,9,11,13,15] Gyr and three different
metallicities: $[Fe/H]=-0.3,~0.0,~+0.3$. Models of increasing age are
marked with symbols of increasing size, the direction of the increase
is shown by the ``Age'' arrow. Also indicated by the dashed arrow is
the direction of increasing metallicity [Fe/H]. Symbols for
observational data are the same of Fig. \ref{mapparent}, NGC\,404 is
marked with a full triangle. The case of NGC\,5557 is not considered
since no group distance is known for this galaxy (Table
\ref{tab6}). The median error bars are indicated lower right in the panel. 
{\it Right panel} - The same as before, but for BVA01 models, for ages
ranging from 5 Gyr to 17.8 (increasing in steps of 12\%) and
metallicities: $[Fe/H]=-0.4,~0.0,~+0.2$.
\label{models}}
\end{figure}

On the other hand, considering group distances -- which are expected
to give a better estimation of the galaxy DM by averaging over the
individual distance uncertainties -- an interesting dichotomy appears
to emerge. The galaxies NGC\,3258, NGC\,3268, NGC\,4696, and
NGC\,5322, that is four among the farthest objects of the sample, lie
on a fainter $\bar{M}_I$ vs. $(B{-}I)_0$ sequence than the galaxies
NGC\,1344 and NGC\,1407 (and possibly NGC\,404), the nearest objects
of the sample (panels $(e), (f)$).  This behavior could be due to the
presence of some kind of bias in the DM estimated for the farthest
objects or, more generally, to some kind of bias in the DM related to
the distance of the objects\footnote{However, we stress the similarity
of the plots in panels $(e)$ and $(f)$ of Fig. \ref{mabsolute}, the
difference $(DM_{FP}^{Group}-DM_{SBF}^{Group})$ being on average $\sim
0.02$, at most $\sim0.06$. Consequently, if it is a bias, it acts very
similarly with both distance indicators.}. For example, by inspecting
the quality parameters $Q$, and $PD$ reported by T01 for their SBF
measurements, we find $Q/PD\geq 2.7$ for all the nearest galaxies,
while it is $Q/PD \leq 0.7$ for the distant ones\footnote{ Such result
holds independently of the fact that we consider single galaxy or
average group properties. In both cases we find the limits quoted in
text. Specifically, we find ($Q\geq 3.8$, $PD\leq 1.4$) for the
nearest galaxies, and ($Q\leq 1.4$, $PD\geq 2.1$) for the others.  For
the definition of the $Q$, and $PD$ parameters see T01. Briefly:
reliable measurements should have high values of $Q$, and low $PD$.}.
Thus, we adopt the $IRAS$ group distances, that is:
$\bar{M}_I=(\bar{m}_I-DM_{IRAS}^{Group})$.
Note, however, that even assuming the IRAS distances there is still
some dichotomy, in the sense that NGC\,1344, and NGC\,4696 still
appear offset in this plane with respect to the other galaxies
(Fig. \ref{mabsolute}, panel $(d)$). This suggests that the
explanation may be a stellar population effect, rather than purely a
relative distance error.  In the following we will address this issue
by comparing observational SBF and color data with theoretical
predictions.
Fig. \ref{models} (left panel) compares SBF and color data from the
present work with our new SBF models which are an upgraded version of
the Paper I models for old metal-rich stellar populations (Cantiello
2004; Paper II); the right panel shows the models from BVA01.  These
two sets of models provide useful independent comparisons because they
use different input stellar tracks, transformations to the empirical
plane, and prescriptions for the later evolutionary phases.  However,
the implications of these independent models for the data are at least
qualitatively similar.

As a general remark, we note that the innermost annuli of the galaxies
have redder $(B{-}I)_0$ colors and fainter SBF magnitudes, with
respect to the outer annuli.  As shown also by the arrows in
Fig. \ref{models}, simple age variations are expected to give
shallower $\bar{M}_I$ vs. $(B{-}I)_0$ gradients with respect to
chemical composition variation.
Interestingly, the succession of the annular color-SBF measurements
for individual galaxies in Fig. \ref{models} follows preferentially
the lines of increasing metallicity toward the center of the galaxy,
with smaller changes in age.  These results agree with the conclusions
of many other authors \citep[e.g.,][]{trager00i}. More specifically,
the results for NGC\,1407, the Antlia group galaxies, and NGC\,5322
agree with this general picture of the SBF and colour variations being
driven by $[Fe/H]$ gradients, but for NGC\,1344 the models indicate
the presence of a younger stellar population in the outward from the
center of the galaxy. As noted above, this behavior, which results in
NGC\,1344 having the smallest value for the SBF--color slope $\beta$,
is consistent with the idea that NGC\,1344 has undergone a merger or
accretion event in the relatively recent past.

From inspection of both panels in Fig. \ref{models}, it appears that
all the data for the giant galaxies are generally consistent with 
intermediate-age to old, metal-rich ($t \geq 5$~Gyr, $[Fe/H] \geq -0.3$) stellar
populations. For NGC\,404, the SBF--color data indicate
a population of half solar metallicity and age $\la 5$~Gyr,
consistent with the results from the data/models comparison
by \citet{jensen03}, and the findings of \citet{schmidt90},
though those authors analyze a smaller region.
However, we caution that any detailed conclusions regarding absolute
metallicities and ages depend on the assumed distance moduli.  For
instance, while the apparent dichotomy between NGC\,1344 and the other
galaxies can be explained by an offset of a factor of $\sim\,$3 in
mean metallicity, this would require exclusively large ages of 9--13
Gyr in this galaxy.  A younger age and more metal-rich population is
found for this galaxy from optical--near-IR SBF colors
(\citealt{jensen03}; Paper I) and line-strength analyses
\citep{kuntschner02}, and would be more consistent with the
indications of fairly recent merging.

\begin{figure}[t]
\epsscale{1.0}
\plottwo{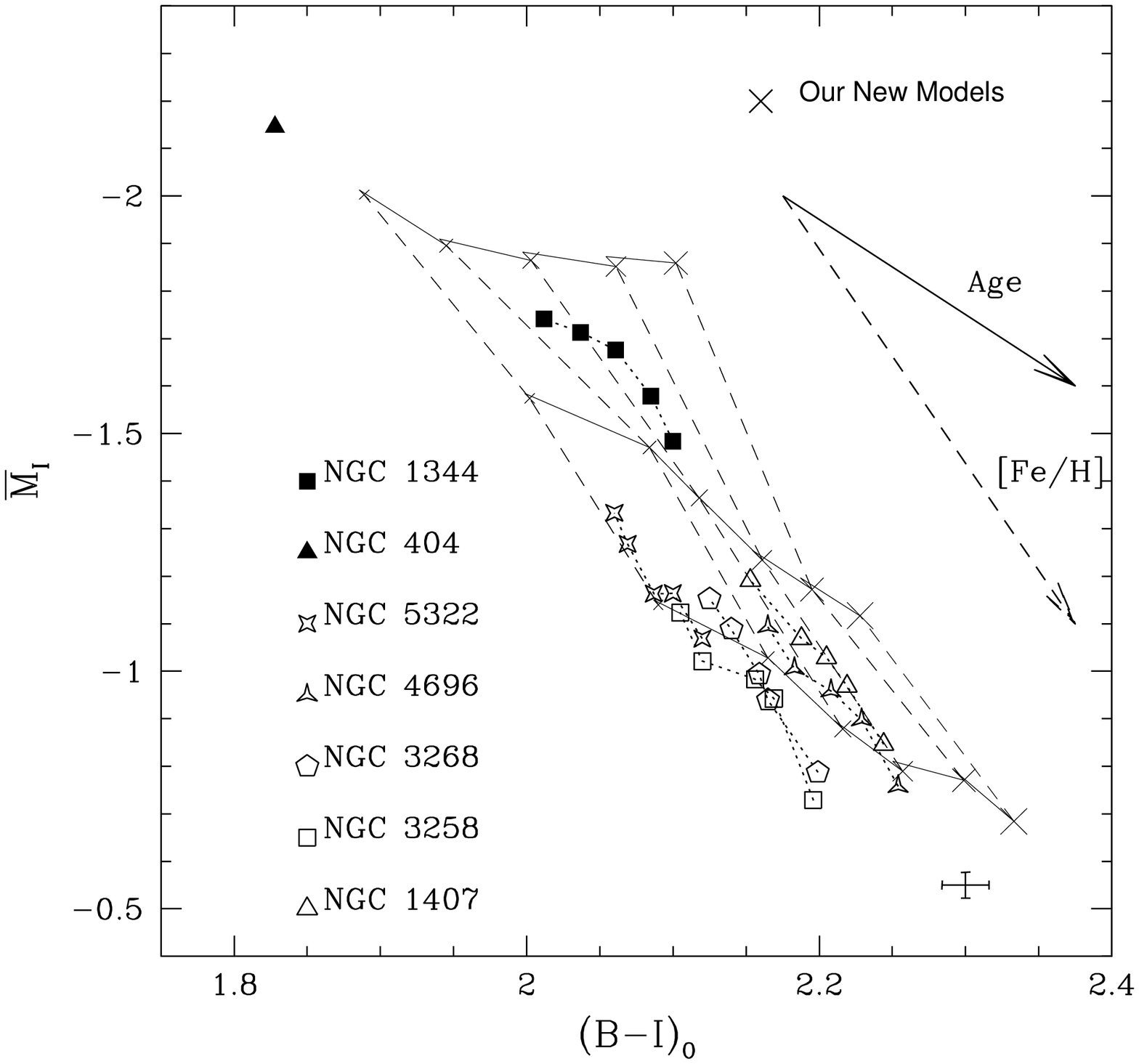}{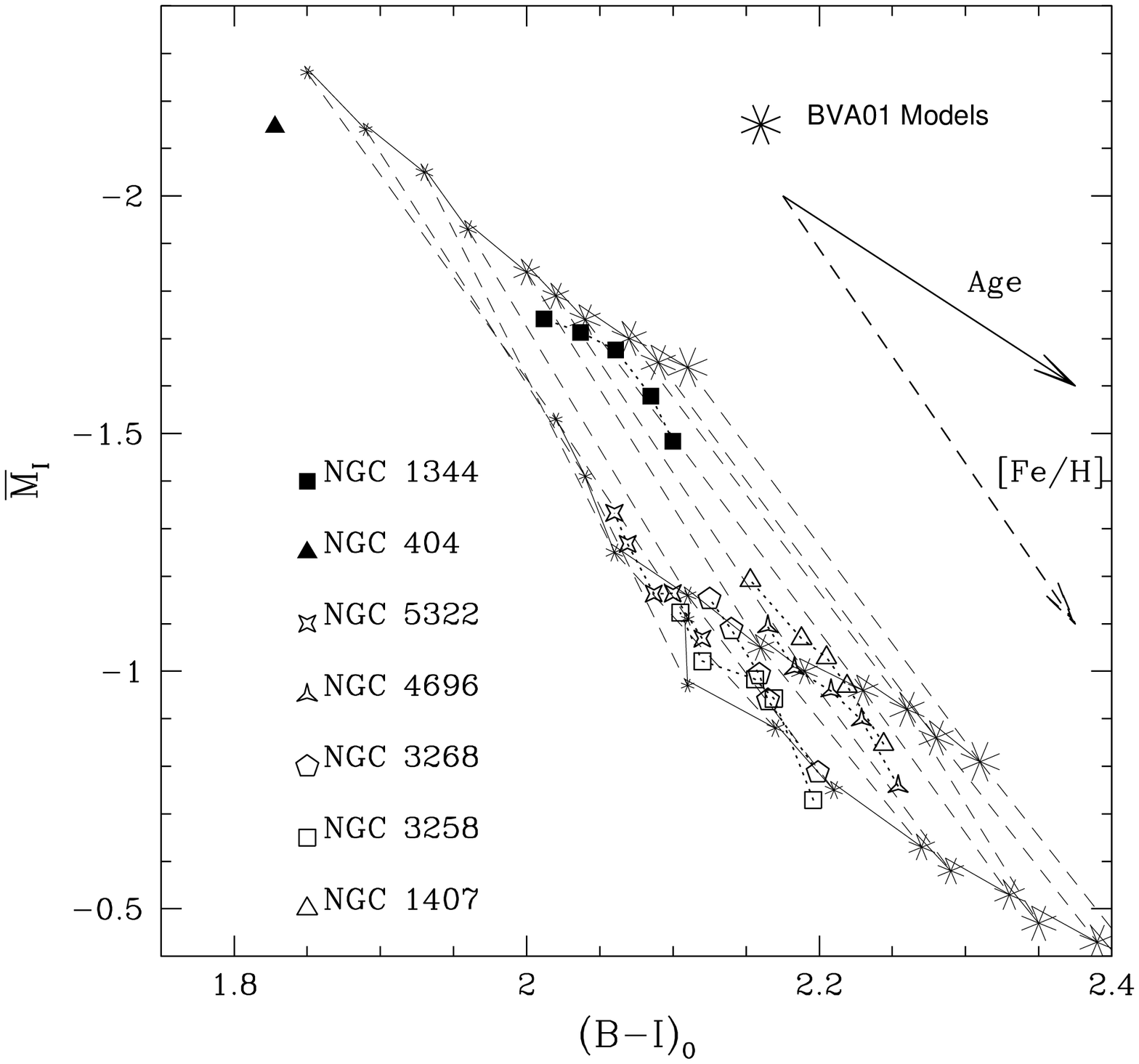}
\figcaption{As Fig. \ref{models} but for average IRAS+FP group distances (see text).
\label{tesi6}}
\end{figure}

To help illustrate the uncertainty further, we present in Figure
\ref{tesi6} a comparison of the data with models assuming the average
of the IRAS and FP group distances.  In this case, the galaxy
NGC\,1407 is confirmed to host an old stellar population of nearly
constant $t \sim 11$~Gyr, showing a significant increase of metal
content in the innermost regions, while the implied age for NGC\,4696
is reduced by nearly a factor of two. NGC\,1344 is predicted to be
dominated by a [Fe/H]${\,\gsim\,}{-}0.3$ old population, but the
disagreement with the results from the near-IR SBF and line-strength
data is substantially reduced.
The galaxies NGC\,5322, NGC\,3258 and NGC\,3268, which are on average more
massive than NGC\,1407 and NGC\,1344, remain consistent with an
intermediate-to-old ($t\sim 5$--9 Gyr), super-solar metallicity
stellar system.   

\begin{figure}[t]
\epsscale{1.}
\plotone{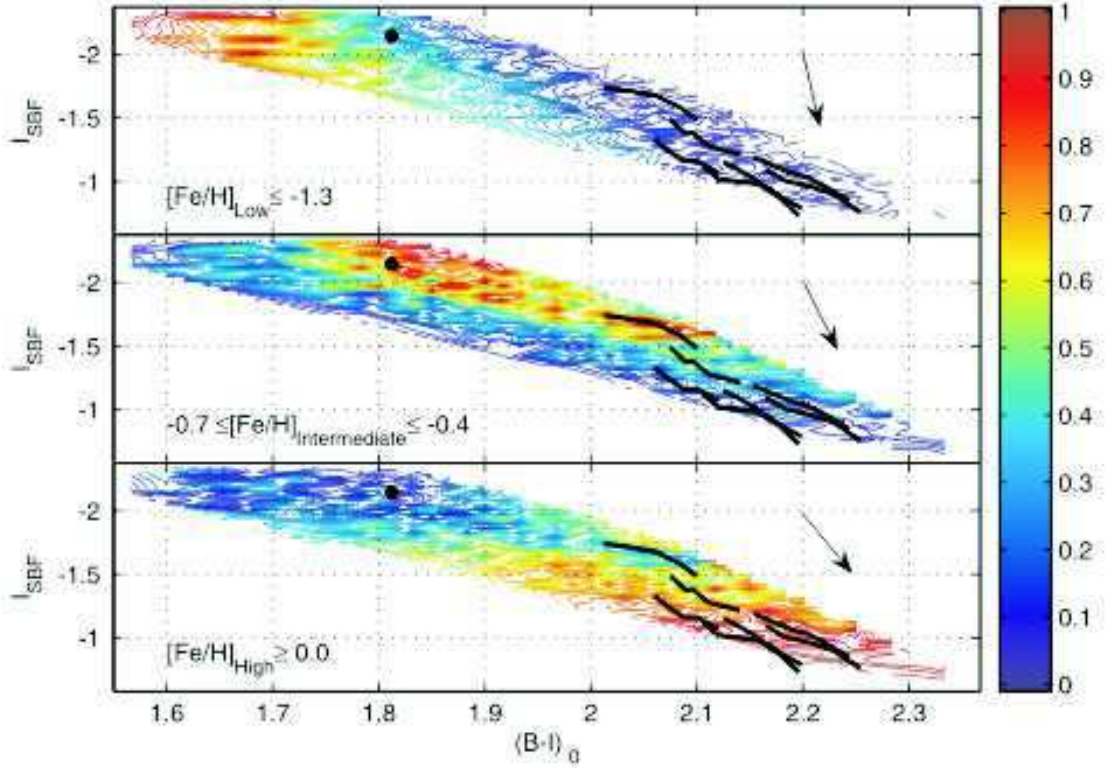}
\figcaption{Composite stellar population models from BVA01, compared to
observational data from this paper (solid black lines). For sake of
clearness we have splitted the BVA01 composite models in three
metallicity classes: low (upper panel), intermediate (middle panel),
and high (lower panel) $[Fe/H]$ populations. The bar on the right
shows the scale of colors associated to the relative weights (in
\%) of each one $[Fe/H]$-component in the model. That is, redder (bluer) 
colors indicate that the component is dominant (negligible) in the
composite model. The spread in age affects mainly the $[Fe/H]_{\rm High}$
models which are computed for $4~Gyr\leq t \leq 18 Gyr$, the
$[Fe/H]_{\rm Low}$, $[Fe/H]_{\rm Intermediate}$ models are computed with ages ranging
from 14 to 18 Gyr, and from 9 to 18 Gyr, respectively. The arrow in
each panel indicates the direction of models with increasing age; the
different slopes of arrows depict the different amplitudes of the age
interval chosen for each metallicity class, i.e., steeper arrows
correspond to narrower age intervals. See text for some more details
on the figure.
\label{tesi7}}
\end{figure}

As a final illustration of how such data could eventually be used to
study stellar populations in more detail, we compare in
Fig.~\ref{tesi7} our observational data with the composite stellar
population models presented by BVA01, assuming again the FP+IRAS
average group distances. In addition to the $[Fe/H]$ interval
considered in the preceding models, these composite models are
computed by taking into account also a stellar component with
metallicity in the range $-1.7\leq[Fe/H]\leq-0.4$. We split the BVA01
models into three metallicity classes, regardless of age: low (upper
panel, $[Fe/H]_{\rm Low}$), intermediate (middle panel, $[Fe/H]_{\rm
Intermediate}$), and high (lower panel, $[Fe/H]_{\rm High}$) metallicity
populations.  The spread in model age mainly comes in for the
$[Fe/H]_{\rm High}$ component of the BVA01 models, which is computed
by those authors for $4~Gyr\leq t \leq 18~Gyr$, while the $[Fe/H]_{\rm
Low}$, and $[Fe/H]_{\rm Intermediate}$ components are computed with ages
ranging from 14 to 18 Gyr, and from 9 to 18 Gyr, respectively.

The color coding (see the color bar on the right of the figure) shows
the percentage of each $[Fe/H]$-component contributing to the
composite population at a given point in the $\bar{M}_I$
vs. $(B{-}I)_0$ plane.  For instance, red indicates that the
particular region of the diagram would be populated by galaxies of
mean metallicity equal to value for that particular panel, while blue
indicates a negligible contribution from this metallicity component.
As an example of the possible mixtures of populations able to
reproduce the NGC\,404 data in this diagram (black dot in
Fig. \ref{tesi7}), these composite models give a dominant stellar
component intermediate in $[Fe/H]$ ($\sim$65\% up to $\sim$90\%), with
a substantial contribution of low metallicity stars ($\sim$15\% to
$\sim$40\%), and a very low percentage of high $[Fe/H]$ stars
($\sim$0\% to $\sim$25\%).
However, the mix of populations needed to reproduce a given point in
the $\bar{M}_I$  vs. $(B{-}I)_0$ plane is not unique, and additional
observables (e.g., optical and near-IR SBF colors) are needed to
achieve firm, distance-independent constraints on composite stellar
population modeling.

\section{Summary}

We have presented a photometric study of eight ellipticals imaged with
the ACS camera on board of HST. Classical broad-band $(B{-}I)_0$ colors,
$B$-, $I$-band magnitudes, and $I$-band SBF magnitudes are determined for
each galaxy.

Given the exceptional resolution of the ACS camera, we succeeded in
measuring SBF and integrated color gradients within galaxies.  This
has allowed us to study the slope of the $\bar{M}_I$ vs. $(B{-}I)_0$
relation in a quite different way compared to earlier applications.
Previous studies derived the slope analyzing multiple galaxies within
the same group, that is, assuming the cluster depth to be negligible
compared to the distance from the observer.  The data analysis
presented in this work does not rely on any assumption about group or
cluster scatter, and shows that, by using galaxy internal SBF
gradients, the uncertainty in the $\bar{M}_I$ vs. $(B{-}I)_0$ slope
can be substantially reduced.

We derive the zeropoint of the empirical $\bar{M}_I$ vs. $(B{-}I)_0$
relation using our measurements for NGC\,1344 (a member of the Fornax
cluster) combined with the known Cepheid distance to Fornax, and thus
present the first full empirical calibration of the $I$-band SBF method
for the ACS instrument.

Concerning the use of SBF as tracer of stellar population properties,
our results substantially confirm the generally well established view
of ellipticals as complex objects, typically dominated by an old,
[Fe/H]$\geq-0.3$ stellar population.  SBF radial gradients are
introduced as a new inquiry tool.  In particular, inspection of the
SBF and color gradients in comparison to models suggests that radial
changes are preferentially due to chemical composition changes, more
than to age, with the outer regions of galaxies being more metal poor
than the innermost regions. Assuming the distance of galaxies from
other indicators, we show that the theory/observations comparison in
the $\bar{M}_I$ vs. $(B{-}I)_0$ plane can help to disentangle the
age/metallicity degeneracy.
While the limited data sample precludes any firm conclusions, we 
consider the results of this study promising for future investigations
of galaxy enrichment histories incorporating SBF and color gradients.

\acknowledgements
ACS was developed under NASA contract NAS 5-32865. Financial support
for this work was provided by MIUR--Cofin 2003. This work made use of
computational resources granted by the Consorzio di Ricerca del Gran
Sasso according to the Progetto 6 {\it ``Calcolo Evoluto e sue
Applicazioni (RSV6)''}--Cluster C11/B.  JPB and SM have been supported
by NASA grant NAG5-7697.  We thank John Tonry for helpful discussions.
We thank the referee for her/his comments useful for the improvement
of this paper.

\appendix
\section{Transforming $(B{-}I)_0$ into $(V{-}I)_0$, and vice versa}
The main body of the data used for this work consisted of F814W, and
F435W long exposures of low redshift bright ellipticals.  Since for
the other two galaxies -- NGC\,404 and NGC\,1344 -- no F435W data are
available, in order to make an homogeneous comparison with the other
galaxies we needed to derive their $(B{-}I)_0$ color from the
available V and I band data.

For NGC\,1344, an exposure in the F606W filter is available from the
same proposal, thus we determined the $(F606W{-}F814W)_0$ radial
profile for this galaxy, then $(F606W{-}F814W)_0$ was transformed into
the standard $(V{-}I)_0$ by using the S05 transformations, finally
$(V{-}I)_0$ was in turn transformed to $(B{-}I)_0$.
For NGC\,404 we only had an F814W image, consequently the $(B{-}I)_0$
color was derived from the $(V{-}I)_0$ measurement for this galaxy given
by T01.

In both cases the $(V{-}I)_0$--to--$(B{-}I)_0$ transformation
equations are derived from the upgraded version of the \citet{cantiello03} 
stellar populations models, by fitting a straight line to the
$(B{-}I)_0$-$(V{-}I)_0$ theoretical predictions for simple stellar
populations in the age range $5~{\rm Gyr} \leq t \leq 15$ Gyr, and in the
metallicity range $-0.3\leq [Fe/H] \leq 0.3$.
The fitted equations are:
\begin{equation}
(B-I)_{Transf}=(2.15\pm0.02)\cdot(V-I)_0+(-0.45\pm0.02) .
\label{eqbivi}
\end{equation}

To check the reliability of this transformation, the $(V{-}I)_0$
measurements from T01 are transformed to $(B{-}I)_0$ applying
eq.~(\ref{eqbivi}), the resulting $(B-I)_{Transf.}$ is then compared
to $(B{-}I)_0$ from this paper.

\begin{figure}[t]
\epsscale{0.75}
\plotone{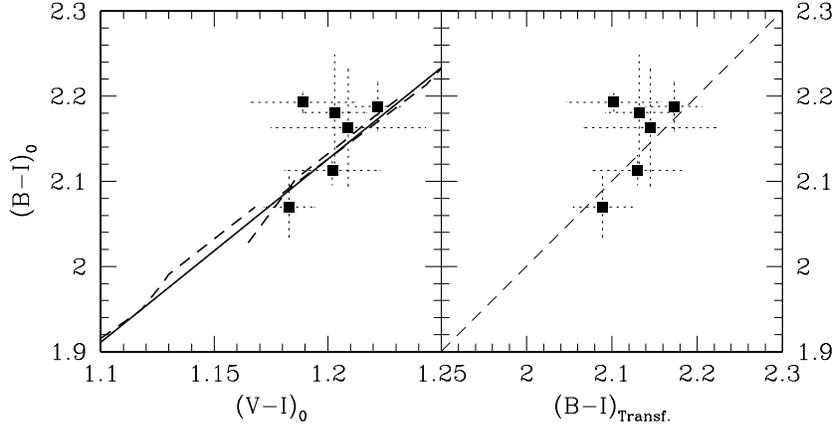}
\figcaption{{\it Left panel } - Distribution of theoretical prediction for
old ($t\geq 5~Gyr$), $[Fe/H]\geq -0.3$ stellar populations in the
$(B{-}I)_0$-$(V{-}I)_0$ plane. Dashed lines are upgraded version of
\citet{cantiello03} models of simple stellar populations, also the
best fit line is plotted. Full dots refer to $(B{-}I)_0$ measurements
from this work, and to $(V{-}I)_0$ data from T01 (data measured within
the same galaxy's region).  {\it Right panel } - $(B-I)_{Transf}$ color
derived applying eq.~(\ref{eqbivi}) to the $(V{-}I)_0$ measured by T01,
versus $(B{-}I)_0$ measured in this work.  The dashed line is not a fit
of the data.
\label{bivi}}
\end{figure}

The result of the comparison can be found in Tab. \ref{tab7}, and
Fig. \ref{bivi}.  Quoted in the table there are: (1) the galaxy name,
(2) average radius of the annulus contributing to color measurement,
the ratio of the innermost to average radius, the ratio of the
outermost to average radius contributing to the mean color; (3)
the measured $(B{-}I)_0$ color; (4) $(V{-}I)_0$ color from by T01; (5)
$(B-I)_{Transf.}$ color obtained applying eq.~(\ref{eqbivi}).

For sake of completeness we report here the $(B{-}I)_0$-$(V{-}I)_0$
color transformations as predicted by using the BVA01 and the
\citet[][BC03 - their Padua 1994 isochrones and Salpeter IMF]{bc03}
models for Simple Stellar Populations (only populations with $5~{\rm
Gyr} \leq t \leq 15$ and $-0.4\leq [Fe/H] \leq 0.4$ are considered):

\begin{equation}
(B-I)_0^{BVA01}=(2.06\pm0.02)\cdot(V-I)_0+(-0.32\pm0.03),
\label{eqbivi2}
\end{equation}
\begin{equation}
(B-I)_0^{BC03}=(2.09\pm0.02)\cdot(V-I)_0+(-0.39\pm0.03).
\label{eqbivi3}
\end{equation}

At the fiducial color $(V{-}I)_0=1.15$, the three different
transformation equations (\ref{eqbivi}, \ref{eqbivi2}, \ref{eqbivi3})
predict $(B{-}I)_0$ color equal to 2.02$\pm$0.03, 2.06$\pm$0.04,
2.02$\pm$0.04 (respectively: our, BVA01, and BC03 models).  However,
even such small variations can determine non-negligible differences in
some of the applications presented in this work. As an example, the
transformation of eq.~(\ref{eqcal}) [\S\,4.1, $\bar{M}_I$
vs. $(B{-}I)_0$] into eq.~(\ref{eqtrans}) [$\bar{M}_I$
vs. $(V{-}I)_0$] changes significantly according to the set of models
that is used:
\begin{equation}
\bar{M}_I^{BVA01}=(-1.6 \pm 0.2) + (6.2 \pm 0.6) \cdot[(V-I)_0-1.15],
\label{eqtrans2}
\end{equation}
and
\begin{equation}
\bar{M}_I^{BC03}=(-1.6 \pm 0.3) + (6.3 \pm 0.6) \cdot[(V-I)_0-1.15].
\label{eqtrans3}
\end{equation}
Note that results from eqs. \ref{eqtrans2}, \ref{eqtrans3}, and
\ref{eqtrans} still agree with each other at the $1~\sigma$ level.
Interestingly, the zeropoints of equations \ref{eqtrans},
\ref{eqtrans2}, and \ref{eqtrans3} nicely reproduce the $\bar{M}_I$
vs. $(V{-}I)_0$ zeropoint derived by \citet{tonry00} by using group
distances (as a matter of fact, combining all models we find
$\bar{M}_I^{(V{-}I)_0=1.15}=-1.6\pm0.1$, to be compared with the
observationally derived $-1.61\pm0.03$).

\newpage

%%%%%%%%%%%%%%%%%%%%%%%%%%%%%%%%%%%%%%%%%%%%%%%%%%%%%%%%%%%%%%%%%%%%
%%%%%%%%%%%%%%%%%FIGURES%%%FIGURES%%%%%%%%%%%%%%%%%%%%%%%%%%%%%%%%%%
%%%%%%%%%%%%%%%%%%%%%%%%%%%%%%%%%%%%%%%%%%%%%%%%%%%%%%%%%%%%%%%%%%%%

\begin{deluxetable}{lcccccccccc}
\tablewidth{0pt}
\tabletypesize{\scriptsize}
\tablecaption{Observational Data\label{tbl-1}}
\startdata
\hline
\multicolumn{1}{c}{Galaxy} & \multicolumn{1}{c}{Group} & \multicolumn{1}{c}{R.A.} &
\multicolumn{1}{c}{Decl.} & \multicolumn{1}{c}{$v_{cmb}$} & \multicolumn{1}{c}{T} &
\multicolumn{1}{c}{$A_B$} & \multicolumn{1}{c}{Prog.} & \multicolumn{1}{c}{Exp. time} &
\multicolumn{1}{c}{Other ACS} & \multicolumn{1}{c}{Exp.} \\
\multicolumn{1}{c}{} & \multicolumn{1}{c}{(group \#)} & \multicolumn{1}{c}{} &
\multicolumn{1}{c}{} & \multicolumn{1}{c}{($km/s$)} & \multicolumn{1}{c}{} &
\multicolumn{1}{c}{} & \multicolumn{1}{c}{ID\#} & \multicolumn{1}{c}{F814W ($s$)} &
\multicolumn{1}{c}{Filter} & \multicolumn{1}{c}{time ($s$)} \\
\hline
NGC 1407 & Eridanus (32) & 55.052  & -18.581 & 1627 & -5  & 0.297 & 9427 & 680  & F435W   & 1500    \\
NGC 3258 & Antlia   (46) & 157.226 & -35.606 & 3129 & -5  & 0.363 & 9427 & 2280 & F435W   & 5360    \\
NGC 3268 & Antlia   (46) & 157.503 & -35.325 & 3084 & -5  & 0.444 & 9427 & 2280 & F435W   & 5360    \\
NGC 4696 & Centaurus (58) & 192.208 & -41.311 & 3248 & -4  & 0.489 & 9427 & 2320 & F435W   & 5440    \\
NGC 5322 & NGC 5322 (254) & 207.315 & 60.191  & 1916 & -5  & 0.061 & 9427 & 820  & F435W   & 3390    \\
NGC 5557 & \nodata  (228) & 214.605 & 36.494  & 3433 & -5  & 0.025 & 9427 & 2400 & F435W   & 5260    \\
\hline
NGC 404 & \nodata (0)  & 17.363  & 35.718  & -332 & -3  & 0.253 & 9293 & 700  & \nodata & \nodata \\
NGC 1344 & Fornax (31)   & 52.080  & -31.068 & 1086 & -5  & 0.077 & 9399 & 960  & F606W   & 1062    \\
\enddata
\label{tab1}
\end{deluxetable}

\begin{deluxetable}{lc}
\tabletypesize{\scriptsize}
\tablecaption{Parameters used for the Detection, Background estimation,
and Deblending with SExtractor v2.2.2}
\tablewidth{0pt}
\tablehead{
\colhead{Parameter} &
\colhead{Value} }
\startdata
BACK\_SIZE      &  128      \cr
BACK\_FILTERSIZE        &  5        \cr
CLEAN               &  Y        \cr
CLEAN\_PARAM        &  1.5      \cr
DETECT\_MINAREA         &  5        \cr
DETECT\_THRESH          &  1.5      \cr
DEBLEND\_NTHRESH    &  8        \cr
DEBLEND\_MINCONT    &  0.008    \cr
FILTER              &  Y        \cr
FILTER\_NAME        &  gauss\_2.0\_3x3.conv  \cr
PHOT\_APERTURES         &  6        \cr
WEIGHT\_TYPE        &  MAP\_RMS \cr
WEIGHT\_THRESH          &  0, 1.0e30\cr
\hline
BACK\_SIZE\tablenotemark{a}             &  25   \cr
BACK\_FILTERSIZE\tablenotemark{a}   &  3
\enddata
\tablenotetext{a}{Background Parameters adopted for the first run of SExtractor
to obtain the large-scale deviations map}
\label{tab2}
\end{deluxetable}

\begin{deluxetable}{cccccc}
\tabletypesize{\footnotesize}
\tablecaption{Color \& SBF measured in each annulus for the whole sample of galaxies. All data are extinction corrected.}
\tablewidth{0pt}
\tablehead{\colhead{Annulus} &  \colhead{$\langle r (arcsec) \rangle$}  & \colhead{$e$\tablenotemark{a}}& \colhead{$(B-I)_0$} &\colhead{$\bar{m}_I^{uncorr.}$} &\colhead{$\bar{m}_I$}}
\startdata
\multicolumn{6}{c}{NGC1407} \\
\hline
1 &  10  & 0.05 &  2.244   0.003 &  31.01 & 31.15   0.05  \\
2 &  16  & 0.05 &  2.219   0.006 &  30.94 & 31.02   0.03  \\ 
3 &  22  & 0.05 &  2.205   0.009 &  30.90 & 30.97   0.03  \\ 
4 &  32  & 0.04 &  2.188   0.015 &  30.86 & 30.92   0.02  \\
5 &  43  & 0.04 &  2.153   0.025 &  30.74 & 30.80   0.02  \\   
\hline
\multicolumn{6}{c}{NGC3258} \\
\hline
1 & 9  & 0.14 &   2.196   0.004 &  31.99 & 32.08   0.04 \\
2 & 15 & 0.10 &   2.169   0.010 &  31.80 & 31.86   0.03 \\
3 & 22 & 0.08 &   2.156   0.019 &  31.75 & 31.82   0.03 \\ 
4 & 31 & 0.08 &   2.120   0.035 &  31.70 & 31.78   0.02 \\
5 & 42 & 0.10 &   2.105   0.060 &  31.58 & 31.68   0.03 \\
\hline
\multicolumn{6}{c}{NGC3268} \\
\hline
1 &  9  & 0.19 & 2.199   0.005 &  31.93 & 32.02   0.03  \\
2 &  15 & 0.20 & 2.165   0.011 &  31.80 & 31.87   0.03  \\
3 &  20 & 0.21 & 2.159   0.018 &  31.75 & 31.81   0.02  \\
4 &  29 & 0.21 & 2.140   0.030 &  31.66 & 31.71   0.02  \\ 
5 &  40 & 0.22 & 2.125   0.050 &  31.56 & 31.65   0.02  \\
\hline
\multicolumn{6}{c}{NGC4696} \\
\hline
1 & 11 & 0.14 &  2.254   0.004 & 31.97 & 32.16   0.06  \\
2 & 15 & 0.15 &  2.229   0.006 & 31.91 & 32.02   0.04  \\
3 & 21 & 0.16 &  2.208   0.008 & 31.88 & 31.96   0.03  \\
4 & 30 & 0.17 &  2.183   0.013 & 31.83 & 31.91   0.03  \\
5 & 38 & 0.19 &  2.165   0.018 & 31.75 & 31.82   0.02  \\
\hline
\multicolumn{6}{c}{NGC5322} \\
\hline
1 & 8  & 0.31 &  2.120   0.002 &   31.19 & 31.38   0.06  \\
2 & 14 & 0.30 &  2.100   0.003 &   31.20 & 31.28   0.03  \\
3 & 19 & 0.31 &  2.087   0.005 &   31.22 & 31.29   0.03  \\
4 & 28 & 0.32 &  2.069   0.009 &   31.13 & 31.18   0.02  \\
5 & 37 & 0.34 &  2.060   0.016 &   31.07 & 31.12   0.02  \\
\hline
\multicolumn{6}{c}{NGC5557} \\
\hline
1 & 9   & 0.19 &  2.141   0.004 & 32.16 & 32.24   0.04 \\
2 & 15  & 0.16 &  2.111   0.009 & 32.11 & 32.17   0.03 \\
3 & 21  & 0.16 &  2.096   0.014 & 32.02 & 32.07   0.02 \\
4 & 30  & 0.15 &  2.089   0.025 & 32.01 & 32.07   0.02 \\
5 & 41  & 0.14 &  2.075   0.051 & 31.89 & 31.96   0.02 \\
\hline
\multicolumn{6}{c}{NGC1344\tablenotemark{b}} \\
\hline
1 & 9  & 0.30 &  2.10   0.02 &   30.04 &   30.17   0.04  \\
2 & 14 & 0.31 &  2.08   0.02 &   30.04 &   30.08   0.02  \\
3 & 19 & 0.34 &  2.06   0.02 &   29.96 &   29.98   0.01  \\
4 & 27 & 0.37 &  2.04   0.02 &   29.94 &   29.95   0.01  \\
5 & 37 & 0.38 &  2.01   0.03 &   29.91 &   29.92   0.01  \\
\hline
\multicolumn{6}{c}{NGC404\tablenotemark{b}} \\
\hline
1 & 10 & 0.08 & 1.81 0.03 & \nodata & 25.42 0.01 \\  
2 & 16 & 0.06 & 1.81 0.03 & \nodata & 25.39 0.01 \\
3 & 22 & 0.07 & 1.81 0.03 & \nodata & 25.43 0.01 \\
4 & 31 & 0.08 & 1.81 0.03 & \nodata & 25.45 0.01 \\
5 & 42 & 0.09 & 1.81 0.03 & \nodata & 25.45 0.01 \\
\enddata
\label{tab3}
\tablenotetext{a}{Average ellipticity ($e=1-b/a$) of the annulus.}
\tablenotetext{b}{$(B{-}I)_0$ colors are derived from $(V{-}I)_0$ data,
see Appendix A.}
\end{deluxetable}

\begin{deluxetable}{lclll}
\tabletypesize{\footnotesize}
\tablecaption{Galaxies relative and absolute distance moduli}
\tablewidth{0pt}
\tablehead{\colhead{Galaxy} & \colhead{$\Delta DM$\tablenotemark{a}} & 
\colhead{$DM_{This\,Work}$\tablenotemark{b}} & \colhead{$\langle DM_{Single}\rangle$\tablenotemark{c}} &
\colhead{$\langle DM_{Group}\rangle$\tablenotemark{d}}}
\startdata
NGC1407     &   0.0            &   32.0 $\pm$ 0.1  &	32.1 $\pm$ 0.15 & 32.01 $\pm$ 0.05 	\\
NGC3258     &   1.0 $\pm$ 0.1  &   33.0 $\pm$ 0.15 &	33.0 $\pm$ 0.15 & 32.8  $\pm$ 0.1   	\\
NGC3268     &   1.0 $\pm$ 0.1  &   33.0 $\pm$ 0.15 &	32.9 $\pm$ 0.15 & 32.8  $\pm$ 0.1	\\
NGC4696     &   1.0 $\pm$ 0.09 &   33.0 $\pm$ 0.1  &	32.8 $\pm$ 0.15 & 32.86 $\pm$ 0.04	\\
NGC5322     &   0.6 $\pm$ 0.08 &   32.6 $\pm$ 0.1  &	32.4 $\pm$ 0.15 & 32.4  $\pm$ 0.1	\\
NGC5557     &   1.4 $\pm$ 0.1  &   33.4 $\pm$ 0.15 &	33.4 $\pm$ 0.1  & \nodata 		\\
NGC1344     &  -0.5 $\pm$ 0.15 &   \nodata  	   &	31.5 $\pm$ 0.3  & 31.53 $\pm$ 0.03	\\
\enddata
\label{tab4}
\tablenotetext{a}{$\Delta DM = DM_{NGC1407}-DM_{galaxy}$}
\tablenotetext{b}{The absolute DM are
derived assuming NGC1344 (Fornax Cluster) as reference, at a
distance of $DM=31.5\pm0.1$.}
\tablenotetext{c}{Distance moduli derived from the weitghted average
of columns 2--4 in Table \ref{tab6}.}
\tablenotetext{d}{Distance moduli derived from the weitghted average
of columns 5, 7, and 8 in Table \ref{tab6}.}
\end{deluxetable}

\begin{deluxetable}{lccc}
\tabletypesize{\footnotesize}
\tablecaption{Slope fitted for single galaxy}
\tablewidth{0pt}
\tablehead{\colhead{Galaxy} & \colhead{$\beta$} & \colhead{$\chi^2$} &\colhead{$q$}}
\startdata
NGC1407     &   3.8 $\pm$ 1.1 &   0.333 &   0.954  \\
NGC3258     &   6.4 $\pm$ 2.0 &   1.014 &   0.798  \\
NGC3268     &   4.7 $\pm$ 1.4 &   0.063 &   0.996  \\
NGC4696     &   3.3 $\pm$ 0.9 &   0.945 &   0.815  \\
NGC5322     &   3.9 $\pm$ 1.0 &   1.568 &   0.667  \\
NGC5557     &   3.5 $\pm$ 1.0 &   0.935 &   0.817  \\
NGC1344     &   3.0 $\pm$ 2.2 &   0.252 &   0.969  \\
\enddata
\label{tab5}
\end{deluxetable}

\begin{deluxetable}{l|ccc|ccccc}
\rotate
\tabletypesize{\footnotesize}
\tablecaption{Distance Moduli assumed to derive the galaxies' SBF absolute magnitudes}
\tablewidth{0pt}
\tablehead{
\multicolumn{1}{l|}{Galaxy} & \multicolumn{1}{c}{$SBF_{T01}$\tablenotemark{a}} & \multicolumn{1}{c}{$FP_{B02}$} &
\multicolumn{1}{c|}{$IRAS_{REF}$} & \multicolumn{1}{c}{$SBF_{T01}$} &
\multicolumn{1}{c}{$N_{SBF}$\tablenotemark{b}} & \multicolumn{1}{c}{$FP_{B02}$} &
\multicolumn{1}{c}{$IRAS_{REF}$} & \multicolumn{1}{c}{$N_{FP+IRAS}$\tablenotemark{b}} \\
\multicolumn{1}{l|}{} & \multicolumn{3}{c|}{Single Distances} & \multicolumn{5}{c}{Group Distances}
}
\startdata
NGC1407 &  32.30 $\pm$ 0.26 & 31.96 $\pm$ 0.41  & 32.04 $\pm$ 0.24 & 32.00 $\pm$ 0.08 &  7 & 32.04 $\pm$ 0.08 & 31.95 $\pm$  0.11 &   5 \\
NGC3258 &  32.53 $\pm$ 0.27 & 33.73 $\pm$ 0.41  & 33.00 $\pm$ 0.19 & 32.64 $\pm$ 0.15 &  3 & 32.63 $\pm$ 0.18 & 32.98 $\pm$  0.14 &   2 \\
NGC3268 &  32.71 $\pm$ 0.25 & 33.34 $\pm$ 0.41  & 32.97 $\pm$ 0.19 & 32.64 $\pm$ 0.15 &  3 & 32.63 $\pm$ 0.18 & 32.98 $\pm$  0.14 &   2 \\
NGC4696 &  32.75 $\pm$ 0.17 & 32.19 $\pm$ 0.41  & 33.12 $\pm$ 0.24 & 32.64 $\pm$ 0.08 &  9 & 32.64 $\pm$ 0.08 & 33.19 $\pm$  0.07 &   9 \\
NGC5322 &  32.47 $\pm$ 0.23 & 31.74 $\pm$ 0.41  & 32.46 $\pm$ 0.20 & 32.29 $\pm$ 0.15 &  4 & 32.34 $\pm$ 0.19 & 32.55 $\pm$  0.13 &   2 \\
NGC5557 &  \nodata    & 32.92 $\pm$ 0.41  & 33.49 $\pm$ 0.11 & \nodata    &\nodata &\nodata &  \nodata    & \nodata \\
\hline
NGC404  &  27.57 $\pm$ 0.10 &  \nodata    & \nodata    & \nodata    &\nodata &\nodata &  \nodata    & \nodata \\
NGC1344 &  31.48 $\pm$ 0.30 &  \nodata    &  \nodata   & 31.49 $\pm$ 0.04 & 26 & 31.55 $\pm$ 0.06 & 31.76 $\pm$  0.10 &   9 \\
\tablenotetext{a}{As discussed in the following sections, the quality parameters
(Q, PD) for SBF data are not equally good for all these galaxies. NGC4696
and NGC3258 are the worst cases, though they are included in the T01
{\bf table of good data (their {\it table.good} table)}. For NGC5557 no SBF data have been published 
up to now.}
\tablenotetext{b}{Number of elements considered to derive the group distance.
Note that some elements of the same group are not common in the SBF-distance and FP-SMAC surveys.
Consequently the group distance are not evaluated with the same dataset. 
For IRAS and FP group distances we adopt the same group definition.}
\enddata
\label{tab6}
\end{deluxetable}

\begin{deluxetable}{lcccc}
\tabletypesize{\footnotesize}
\tablecaption{$(B-I)_0$ to $(V-I)_0$ transformations}
\tablewidth{0pt}
\tablehead
{\colhead{Galaxy} & \colhead{$\langle r\rangle $\tablenotemark{a}}  & \colhead{$(V-I)_0$} & \colhead{$(V-I)_0^{T01}$} & \colhead{$(B-I)_{Transf.}$}}
\startdata
NGC404  &  23 0.0 3.8 & \nodata     &  1.054 $\pm$ 0.011 &    1.812 $\pm$ 0.034 \\
NGC1344 &  27 0.3 2.2 & 1.150\tablenotemark{b} $\pm$ 0.025 &  1.135 $\pm$ 0.011 &   1.986 $\pm$ 0.034 \\
\cutinhead{Galaxy~~~~~~~~~~~~$\langle r \rangle$\tablenotemark{a}~~~~~~~~~~$(B-I)_0$~~~~~~~~$(V-I)_0^{T01}$~~~~~~$(B-I)_{Transf.}$}
NGC1407 &  31 0.2 1.9   &   2.188 $\pm$ 0.029 &   1.232 $\pm$ 0.018 &   2.173 $\pm$ 0.033 \\
NGC3258 &  26 0.3 2.3   &   2.163 $\pm$ 0.069 &   1.220 $\pm$ 0.036 &   2.145 $\pm$ 0.077 \\
NGC3268 &  9  0.5 1.9   &   2.193 $\pm$ 0.012 &   1.235 $\pm$ 0.012 &   2.102 $\pm$ 0.056 \\
NGC4696 &  57 0.3 2.1   &   2.181 $\pm$ 0.067 &   1.229 $\pm$ 0.035 &   2.132 $\pm$ 0.039 \\
NGC5322 &  29 0.4 3.0   &   2.070 $\pm$ 0.036 &   1.174 $\pm$ 0.020 &   2.089 $\pm$ 0.034 \\
NGC5557 &  25 0.2 1.7   &   2.113 $\pm$ 0.017 &   1.195 $\pm$ 0.013 &   2.130 $\pm$ 0.052 \\
\tablenotetext{a}{The sub-columns report: (i) the average radius of the annulus contributing to color
measurement, (ii) the ratio of the innermost to average radius, and (iii) the ratio
of the outermost to average radius contributing to the mean color, respectively.}
\tablenotetext{b}{This $(V-I)_0$ comes from the observed
$(F606W-F814W)_0=0.876 \pm 0.016$, transformed according to the S05 prescriptions.}  
\enddata
\label{tab7}
\end{deluxetable}

\newpage

\bibliographystyle{apj}
\bibliography{cantiello}

\end{document}